# Anti-bullying Adaptive Cruise Control: A proactive right-of-way protection approach

Jia Hu, *Senior Member, IEEE,* Zhexi Lian, Haoran Wang, *Member, IEEE*, Zihan Zhang, Ruoxi Qian, Duo Li, *Senior Member, IEEE*, Jaehyun (Jason) So, Junnian Zheng

***Abstract*— Adaptive Cruise Control (ACC) systems have been widely commercialized in recent years. However, existing ACC systems remain vulnerable to close-range cut-ins, a behavior that resembles "road bullying". To address this issue, this research proposes an Anti-bullying Adaptive Cruise Control (AACC) approach under a vehicle-cloud cooperation scheme. To handle diverse "road bullying" cut-in scenarios smoothly, the proposed approach first leverages an online Inverse Optimal Control (IOC) based algorithm for individual driving style identification in the cloud. Then, based on Stackelberg competition, a game-theoretic-based motion planning framework is presented in which the identified individual driving styles are utilized to formulate cut-in vehicles' reaction functions. By integrating such reaction functions into the ego vehicle's motion planning, the cloud could consider cut-in vehicles' all possible reactions to find optimal right-of-way protection maneuvers for the ego vehicle. To the best of our knowledge, this research is the first to model vehicles' interaction dynamics and develop an interactive planner that adapts cut-in vehicle's various driving styles. Simulation results show that the proposed approach can prevent "road bullying" cut-ins and be adaptive to different cut-in vehicles' driving styles. Compared with the traditional ACC, the proposed approach can improve safety and comfort by up to 79.8% and 20.8%. The driving efficiency has benefits by up to 17.32% in traffic flow. Increasing the penetration rate of cloud-connected EVs improves lane-average speed by up to 1.36 km/h, demonstrating traffic-flow-level benefits. Furthermore, the proposed approach can support real-time field implementation by ensuring less than 50 milliseconds computation time.**

This research is partially supported by National Natural Science Foundation of China (Grant No. 52302412 and 52372317), Yangtze River Delta Science and Technology Innovation Joint Force (No. 2023CSJGG0800), Shanghai Automotive Industry Science and Technology Development Foundation (No. 2404), the Fundamental Research Funds for the Central Universities, Tongji Zhongte Chair Professor Foundation (No. 000000375-2018082), Xiaomi Young Talents Program, Shanghai Sailing Program (No. 23YF1449600). (*Corresponding author: Haoran Wang*)

Jia Hu, Zhexi Lian, and Haoran Wang are with Key Laboratory of Road and Traffic Engineering of the Ministry of Education, Tongji University, No.4800 Cao'an Road, Shanghai, China, 201804. (e-mail: hujia@tongji.edu.cn,zhexi_lian@tongji.edu.cn,wang_haoran@tongji.edu.cn); Zihan Zhang is with Shanghai Motor Vehicle Inspection Certification and Tech Innovation Center Co., LTD, 68 South Yutian Road, Shanghai, P.R.China, 201805. (e-mail: zihanz02@smvic.com.cn); Ruoxi Qian is with University College London, 25 Gordon Street, London, United Kingdom. (e-mail: ruoxi.qian.22@ucl.ac.uk); Duo Li is with Department of Civil and Geospatial Engineering, Newcastle University, Newcastle upon Tyne NE1 7RU, UK. (e-mail: duoli0725@gmail.com); Jaehyun So is with the Department of Transportation System Engineering, Ajou University, Suwon, Gyeonggi 16499, South Korea (e-mail: jso@ajou.ac.kr); Junnian Zheng is with Hyperview Mobility (Shanghai) Co., Ltd. No.488 Anchi Rd, Shanghai, 201805, P.R.China. (e-mail: junnian.zheng@hongjingdrive.com)

## I. INTRODUCTION

### A. *Research motivation*

Adaptive Cruise Control (ACC) system is one of the most broadly commercialized driving assistance functions. More than 50% of newly sold vehicles have been equipped with ACC [1]. It is widely recognized for reducing driver workload and improving comfort [2][3]. However, how to deal with cut-ins from adjacent lanes remains challenging for ACC systems. According to [4][5], close-distance (time to collision less than 1.5 seconds) cut-ins can cause the ACC-activated ego vehicle (EV) to brake hard, leading to reduced mobility [6], a unsafe time headway less than 0.6 s [7], and acceleration oscillations more than 3 m/s$^2$ [8], as illustrated in Figure 1. Such cut-ins account for as much as 20.6% of all cut-ins in Shanghai [9]. This research refers to such cut-ins as "road bullying" cut-ins, a type of behavior which resembles bullying other vehicles on the road. Traffic regulations in the US, EU, Japan, etc. have implicitly forbid such unsafe cut-ins. When faced with such "road bullying" cut-ins, existing ACC systems often have to yield and lose the right-of-way, exhibiting an around 40% of ACC disengagement rate [7][8], showing reluctance to be used [10]. Hence, it is necessary to enhance ACC systems' right-of-way protection capability to prevent "road bullying" cut-ins.

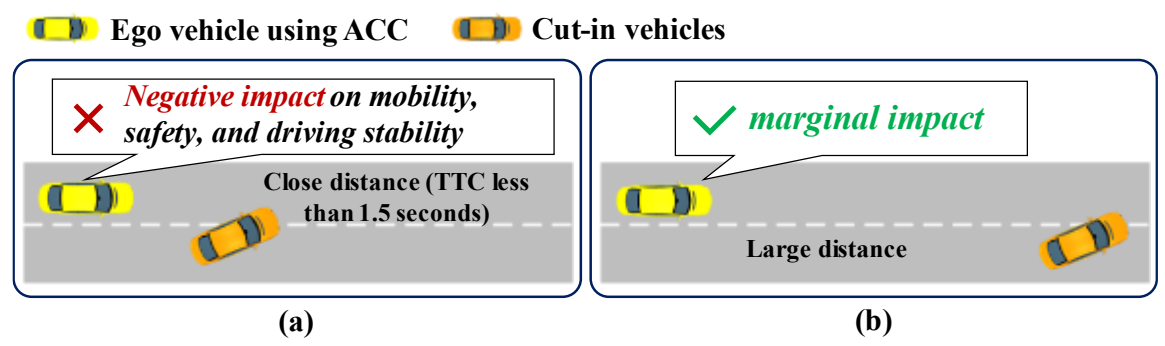

Figure 1 The definition of cut-in types. (a) "road bullying" cut-ins; (b) "no bullying" cut-ins.

### B. *Related work and limitations*

Current mainstream approaches for "road bullying" cut-ins prevention can be divided into three categories: rule-based, model-base, and learning-based approaches.

To prevent "road bullying" cut-ins, the earliest "anti-bullying" ACC systems were developed using rule-based approaches. This type of approach relied on hand-crafted rules to activate different safety-oriented precautions [11][12]. Each precaution was designed to address specific types of cut-in maneuvers [13]. A key contribution was made by Chen et al. [14] presented a Finite State Machine (FSM) to manage state transitions when a vehicle cuts in. He et al. [15] also employed a rule-based FSM framework to handle safety-critical cut-in scenarios. Despite being effective in specific conditions, these rule-based approaches often fail to respond effectively to cut-in maneuvers that the EV has never met before. Given an illustrative example, if another vehicle suddenly cuts in at a very close and never-before-seen distance, the EV cannot avoid a collision because no rules have been developed to respond to the urgent cut-in. This limitation results in poor right-of-way protection capability.

To effectively handle unexpected cut-in maneuvers, model-based approaches have been widely adopted. These approaches address multi-objective optimization problems under safety constraints, ensuring both safety and global optimality during cut-in preventions. However, model-based approaches lack resilience against different driving styles of CVs [27][28][29]. While Liu et al. [30] tested their approach under different cut-in styles, they still cannot conduct person-by-person countermeasure maneuvers. Blindly competing with CVs of unknown driving styles may not only fail to prevent cut-ins but also introduce high safety risks. Furthermore, model-based approaches overlook real-time field implementation capability [31][32]. Mixed-integer and nonlinear problems are time-consuming [23][25], which makes it hard to achieve a control frequency exceeding 10 Hz. Hence, the practicality of the above approaches in real-world applications is limited. To this end, model-based approaches need to be improved.

In some cases, learning-based techniques are incorporated into model-based approaches, further improving adaptability and robustness [34][35][36]. For instance, Bhattacharyya et al. [23] developed a mixed-integer Model Predictive Control (MPC) controller with an online control parameter learning algorithm incorporated for cut-in prevention. However, existing approaches are typically conducted in a "prediction→reaction" fashion, which means predicting the cut-in vehicle (CV)'s intention or trajectory first and responding to the predicted intentions or trajectories secondly. This operation fashion has a critical limitation. This "prediction→reaction" fashion suffer from the passively responding nature. The "passively responding" means the EV must give way to the CV. In other words, EV's optimal maneuver is constrained by CV [24][25]. Lu et al. [24] leveraged predictive control only when cut-in intention was predicted. Similarly, Wei et al. [25] developed a bi-level optimization framework based on game theory, in which the EV's optimal strategy is conditioned on the anticipated actions of CVs. In contrast, skilled human drivers proactively respond to cut-ins by anticipating CV reactions to the EV's motions and adjusting their own trajectory to influence the CV's behavior [26].

To effectively handle the passive limitation using intelligent strategies, recent research has focused more on pure learning-based approaches. One branch of research is Reinforcement Learning (RL). RL learns how to handle "road bullying" cut-ins through trial and error. Zhao et al. [16] explored the potential for RL to manage cut-in events safely. Han et al. [17] further applied deep neural networks combined with RL for cut-in prevention. Liu et al. [30] considered proactive cut-in prevention by tentative acceleration through a rainbow deep Q-network, which is a noteworthy development. Another branch of research is Imitation Learning (IL). IL continuously accumulates driving experience from a wide range of previously encountered cut-in maneuvers [18][19]. By imitating how human drivers respond to these maneuvers, such approaches improve their driving responses over time, becoming adaptive to a wide variety of cut-in scenarios. One representative research is presented by Zhou et al. [20], in which a Long Short-Term Memory (LSTM) based transfer learning framework is leveraged for learning human driving patterns. However, such approaches still have limitations. Firstly, learning-based approaches often lack interpretability and fail to account for vehicle dynamics. As a result, actions produced by such approaches may exceed vehicles' operational capabilities and compromise safety [21]. Secondly, learning-based approaches cannot guarantee the global optimality of driving maneuvers. This occurs because the actions are often heuristic approximations based on previous experiences with inherent randomness, rather than definitive global optimal maneuvers in real-time [22].

### C. Contributions

To overcome the above limitations, this research proposed an **Anti-bullying Adaptive Cruise Control (AACC) approach,** which explores a novel vehicle-cloud cooperation applicable scenario—proactively preventing "road bullying" cut-ins. The AACC algorithm is deployed in the cloud. For every EV which needs to prevent cut-ins, the vehicle-cloud cooperation will be built. Among the cloud computation unit, to handle diverse cut-in behaviors smoothly, an online Inverse Optimal Control (IOC) based algorithm is first leveraged to identify individual driving styles of "road bullying" CVs in real time. Then, this research presents a game-theoretic-based motion planning framework based on Stackelberg competition. In this framework, the identified individual driving styles are utilized to formulate CVs' reaction functions. By integrating such reaction functions into the EV's optimal planning, the cloud could consider CVs' all possible reactions to find optimal right-of-way protection maneuvers for EV. To the best of our knowledge, this research is the first to model vehicles' interaction dynamics and develop an interactive planner that adapts CV's various driving styles. The scientific innovations are detailed as follows:

***a) With the enhanced capability of preventing "road-bullying" cut-ins.*** This anti-bullying capability is achieved by proactively discouraging potential cut-in maneuvers from other vehicles. Specifically, this approach designs a game-theoretic-based motion planning framework based on Stackelberg competition. It works by having the EV take actions to discourage other vehicles' cut-ins under safety-guaranteed conditions. For instance, the EV could accelerate to close the following gap safely when CVs exhibit hesitant behavior, making the gap appear insufficient for a cut-in attempt.

***b) Adaptive to various driving styles of CVs.*** Different driving styles lead to different behavioral characteristics of CVs. The proposed approach leverages an online Inverse Optimal Control (IOC) based algorithm to identify individual driving styles of CVs in real time. Hence, person-by-person countermeasure maneuvers could be produced in response to CVs' individual driving styles. Given an illustrative example, when facing an aggressive CV driver, the EV's accelerating attempt would not suppress the cut-in. At this time, the short-term historical trajectory helps the cloud learn that the CV's driver is aggressive. Updating this CV's characteristic, the cloud would recommend EV yielding to the cut-in, avoiding enraging the CV driver.

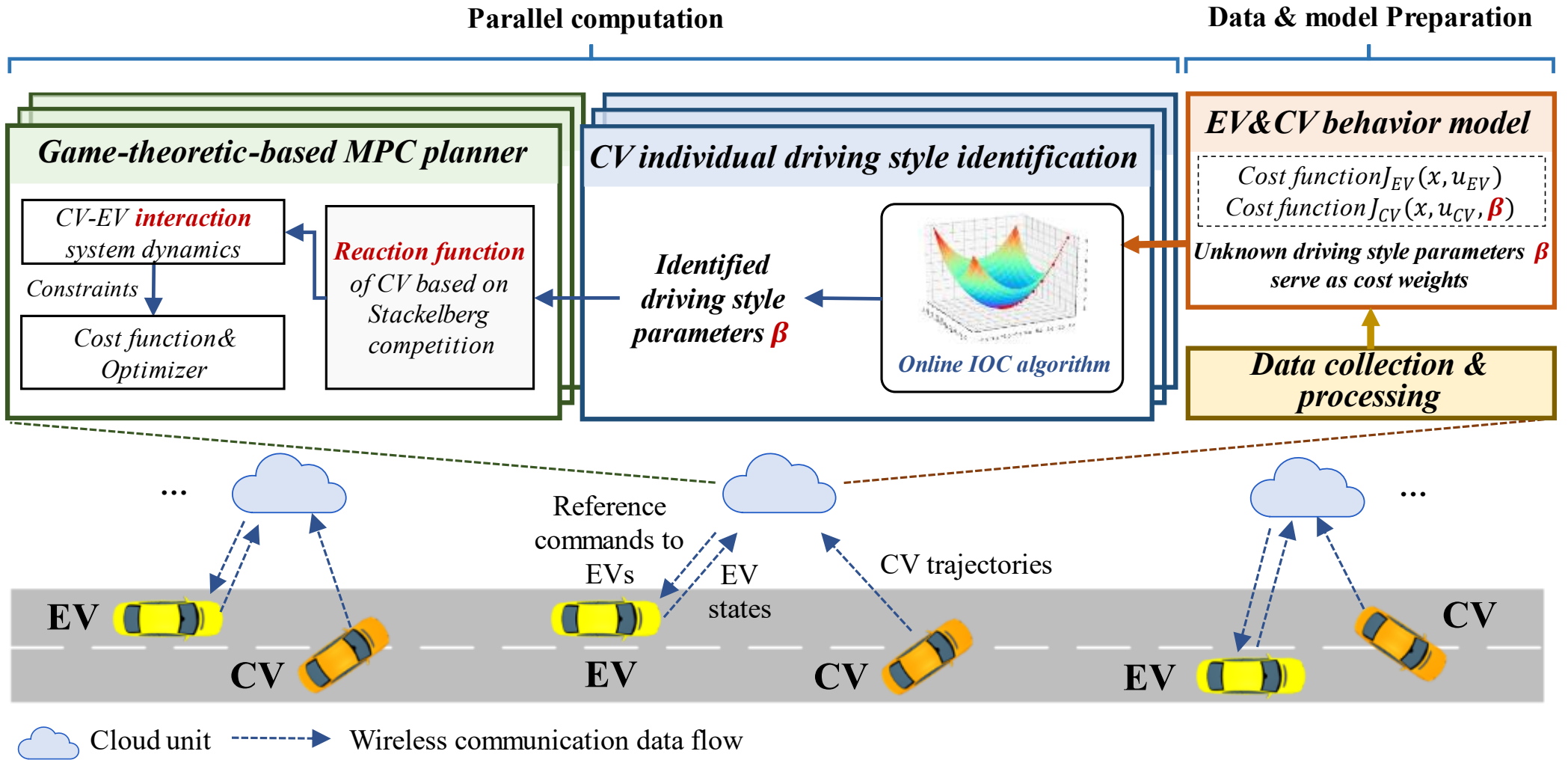

Figure 2 The Overall system structure. The proposed approach is deployed in the cloud unit. It collects EV states and CV trajectory data and sends reference commands back. Among the cloud, in the data & model preparation stage, it processes the data and formulate behavior models. In the parallel computation stage, each CV's driving style parameters can be identified and fed into Game-theoretic-based MPC Planner module to generate reference commands for each EV. $J_{CV}$ and $J_{EV}$ denotes the cost function of the CV and EV behavior model. $x$ , $u_{CV}$ and $u_{CV}$ denote the system state, EV's motion and CV's motion.

***c) Optimal but not unsafe.*** The proposed approach can ensure optimal planning via responding to CVs' all reactions while ensuring safety. The cloud formulates CVs' reaction functions, which describe how CVs react to the EV's motions. The reaction function is derived by solving their behavioral model, which is formulated as an MPC problem. EV's motions are introduced into the CV's dynamics model. Hence, the solution to the MPC problem describes how the motions of the EV affect the driving behavior of the CVs. By integrating the reaction functions into the system dynamics of the EV's optimal planning, the cloud could consider CVs' all possible reactions to find EV's optimal maneuver. Meanwhile, the hard safety constraints can be incorporated to avoid unsafe vehicle interactions.

***d) With real-time field implementation capability.*** To improve the practicality, the proposed approach is formulated as a quadratic programming problem and calculated in the cloud. This formulation ensures computational efficiency in the real-time field implementation.

## II. Problem Formulation

### A. *The vehicle-cloud cooperation scheme*

The vehicle-cloud cooperation scheme is shown in Figure 2, the autonomous vehicles are defined as EVs. The CV in the adjacent lane tends to cut in the EV mandatorily. EV competes with CV to protect right-of-way by preventing "road bullying" cut-ins. The proposed AACC approach is deployed in the cloud unit. It collects EV states and CV trajectory data and sends reference commands back to EVs through wireless communications. This vehicle-cloud cooperation scheme is adopted as i) Compared with onboard perception, the cloud provides access to longer historical trajectories, which can be queried when the EV encounters a cut-in scenario; ii) Obtaining CV trajectories via EV's onboard perception incurs computational overhead and suffers from unavoidable estimation errors, and may also suffer from the moral hazard of privacy leaks. Hence, collecting data and computing in the cloud unit is of great advantages.

### B. *Overall system structure*

The detailed system structure is illustrated in Figure 2. For every EV which needs to prevent cut-ins, the vehicle-cloud cooperation will be built. The first stage in the cloud is data & model preparation: **Module 1.1: cloud data collection and processing.** This module collects EV states and CV trajectories data through wireless communication and processes the data for downstream modeling; **Module 1.2: EV & CV behavior model formulation.** As EV is an autonomous vehicle, its behavior model is built as an optimal control model. As for CV, this research assumes that it follows an MPC behavior model with a predefined cost function, which governs its behavior. The driving style parameters serve as the cost weights in the cost function.

The second stage in the cloud is the parallel computation for multiple pairs of EV and CV. For each pair, the computation consists of two main modules: **Module 2.1: CV individual driving style identification.** This module enables real-time identification of the CVs' individual driving style. Utilizing the processed data and predefined behavior models, an online IOC-based algorithm is employed to infer CVs' driving style parameters. These identified parameters are continuously fed into the planner; **Module 2.2: Game-theoretic-based MPC Planner.** Based on Stackelberg competition**,** this module optimizes person-by-person interdependent driving strategies for EVs to protect right-of-way. Given the identified driving style parameters, the CV's reaction function is obtained through solving CV's behavior model. The Game-theoretic-based MPC (GMPC) planner models the interaction system dynamics between EV and CV, in which CV's reaction function is integrated. The GMPC planner is efficiently solved in the cloud, generating real-time reference control commands and sending back to EV.

### C. Control logic

The proposed approach is formulated under the Partially Observable Markov Decision Process (POMDP) framework, which is well-suited for modeling decision-making under uncertainty [37]. The reason why POMDP adopted is that the unobserved driving style parameters of CV represent the uncertainty of CV's driving behavior. The variety of driving style parameters emulates diverse CV individual driving styles. The POMDP is defined by the tuple $(\mathcal{S}, \mathcal{A}, \mathcal{O}, \mathcal{Z}, \mathcal{T}, \mathcal{R})$. $\mathcal{S}$ is the state space. $\boldsymbol{u}_{EV} \in \mathcal{A}$ is the action space for EV. $\mathcal{O}$ is the observation space, which contains the observed physical state of vehicles, such as position, speed, and acceleration. $\mathcal{Z}$ is the observation model. $\mathcal{T}$ denotes the system state transition model. We model $\mathcal{T}$ as the interaction system dynamics between EV and CV (See Section III. A). $\mathcal{R}$ is the reward function (See Section III. A). In the proposed POMDP framework, EV maintains a belief $\mathcal{B}$ as an estimate of the driving style parameters of CV. The belief $\mathcal{B}$ updates based on the history trajectory data of CV (See section III. B).

### D. Method overview

Section III. A illustrates the data collection and processing, models the CV-EV interaction system dynamics which serves as the state transition model in POMDP, and formulates the behavior model of EV and CV. Section III. B illustrates the online IOC based algorithm which identifies individual driving styles of CVs in real time. Section III.C explains how to build a game-theoretic-based motion planning framework with consideration of CV's reactions.

## III. METHODOLOGY

### A. Cloud data & model preparation

#### 1) Cloud data collection & processing

Through wireless communication technology, we firstly collect the EV state packets and the CV historical trajectories. The raw trajectory stream is first time-aligned and converted into uniformly sampled discrete points to match the downstream IOC identification and GMPC planning that both operate in discrete time. Let the raw CV (or EV) trajectory be $\widetilde{\mathbf{p}_{veh}(t)} = [\widetilde{x_{veh}(t)}, \widetilde{y_{veh}(t)}]$. The cloud resamples it at a fixed step $\Delta t$, producing the discrete trajectory points:

$$\mathbf{p}_{veh,n} = \widetilde{\mathbf{p}_{veh}(t_n)}, t_n = n\Delta t, veh \in \{EV, CV\} \tag{1}$$

Since the collected trajectories may contain measurement noise, the cloud performs a lightweight low-pass filtering on the position sequence via an exponential moving average (EMA):

$$\widetilde{\mathbf{p}_{veh,n}} = (1-\rho)\widetilde{\mathbf{p}_{veh,n-1}} + \rho\mathbf{p}_{veh,n}$$
$$0 < \rho < 1, veh \in \{EV, CV\} \tag{2}$$

Based on the discretized positions, the cloud extracts the speed components by finite differences:

$$\mathbf{v}_{veh,n} = \frac{\widetilde{\mathbf{p}_{veh,n}} - \widetilde{\mathbf{p}_{veh,n-1}}}{\Delta t}, v_{veh,n} = \left|\left|\mathbf{v}_{veh,n}\right|\right|_2 \tag{3a}$$

$$\psi_{veh,n} = \mathrm{atan2}\left(y_{veh,n} - y_{veh,n-1}, x_{veh,n} - x_{veh,n-1}\right) \tag{3b}$$

which provides the speed and yaw angle required by the behavior modeling and identification modules. These processed data then sent to downstream computation in the cloud.

#### 2) CV-EV interaction system dynamics

The CV-EV interaction system dynamics are modeled as the state transition model $\mathcal{T}$. Since CV conducts longitudinal and lateral coupled cut-in maneuvers and EV conducts longitudinal only maneuvers, the system state vector shall include the longitudinal and lateral state of CV and the longitudinal state of EV, defined as:

$$\boldsymbol{x} = (\Delta x, v_{EV}, v_{CV}, y_{CV}, \psi_{CV})^{\mathrm{T}} \tag{4}$$

where $\Delta x$ denotes the longitudinal distance between EV and CV; $v_{EV}$ is the speed of EV; $v_{CV}$ is the speed of CV; $y_{CV}$ represents CV's lateral position and $\psi_{CV}$ represents CV's yaw angle. The reference command for EV is defined as:

$$u_{EV} = a_{EV}, \qquad u_{EV} \in \mathcal{A} \tag{5}$$

where $a_{EV}$ denotes the acceleration of EV.

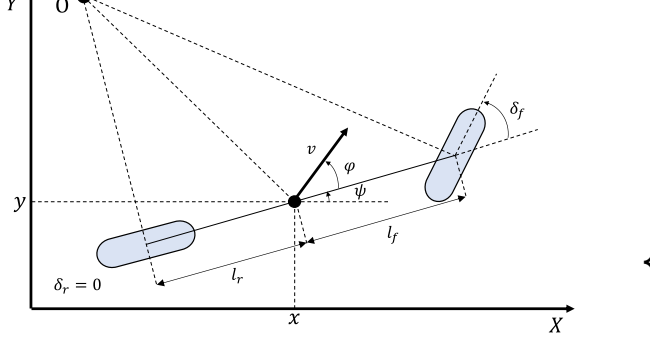

Figure 3 Kinematic bicycle model notation.

$$\begin{cases} \dot{x}_{CV} = v_{CV}\cos\left(\psi_{CV} + \varphi\right) \\ \dot{y}_{CV} = v_{CV}\sin\left(\psi_{CV} + \varphi\right) \\ \dot{\psi}_{CV} = \dfrac{v_{CV}}{l_r}\sin\varphi \\ \dot{v}_{CV} = a_{CV} \\ \varphi = \tan^{-1}\left(\dfrac{l_r}{l_r + l_f}\tan\delta_f\right) \end{cases} \tag{6a}$$

The kinematic model is based on the vehicle kinematic bicycle model [38][39]. The vehicle system is illustrated in Figure 3. $\varphi$ is the angle of the current velocity of the gravity center with respect to the longitudinal axis of the vehicle. For CV, both longitudinal and lateral kinematics need to be considered. The detailed formulation of the model is in equation (6a), where $l_r$ and $l_f$ are distance between vehicle's gravity center and the rear and front axles respectively; $\delta_f$ denotes CV's front steering angle. Based on the small angle assumption of $\varphi$ [40], the kinematic bicycle model, i.e., equation (6a) is linearized as follows:

$$\dot{y}_{CV} = v_{CV}\sin(\psi_{CV} + \varphi) \approx v_{CV}\sin(\psi_{CV}) \approx v_{CV}\psi_{CV}$$
$$\dot{\psi}_{CV} = \frac{v_{CV}}{l_r}\sin\varphi \approx \frac{v_{CV}}{l_r}\tan^{-1}\left(\frac{l_r}{l_r + l_f}\tan\delta_f\right) \approx \frac{v_{CV}}{l_f + l_r}\delta_f \tag{6b}$$

the linearization procedure in equation (6b) mainly depends on the theorem $\sin\theta \approx \theta$ and $\tan\theta \approx \theta$ when $\theta$ is small. This linearization procedure is widely adopted by relevant papers [26][41]. For EV, only longitudinal kinematic need to be considered:

$$\dot{x}_{EV} = v_{EV} \tag{7a}$$

$$\dot{v}_{EV} = a_{EV} \tag{7b}$$

Then, the CV-EV interaction system dynamics (also known as the state transition model $\mathcal{T}$) could be derived as follows:

$$\dot{\boldsymbol{x}} = \boldsymbol{A}\boldsymbol{x} + \boldsymbol{B}u_{EV} + \boldsymbol{C}\boldsymbol{u}_{CV} \tag{8}$$

where,

$$\boldsymbol{A}=\begin{bmatrix}0&1&-1&0&0\\0&0&0&0&0\\0&0&0&0&0\\0&0&0&0&v_{CV}\\0&0&0&0&0\end{bmatrix},\boldsymbol{B}=\begin{bmatrix}0\\1\\0\\0\\0\end{bmatrix},\boldsymbol{C}=\begin{bmatrix}0&0\\0&0\\-1&0\\0&0\\0&\frac{v_{CV}}{l_f+l_r}\end{bmatrix} \tag{9a}$$

$$\boldsymbol{u}_{CV}=(a_{CV},\delta_f)^{\mathrm{T}} \tag{9b}$$

the dimensions of $\boldsymbol{B}$ and $\boldsymbol{C}$ are $5\times 1$ and $5\times 2$ to match the state and input dimensions.

*3) EV & CV behavior model formulation*

The behavior model of EV is the optimal control model, in which the cost function is defined as:

$$J_{EV}=\frac{1}{2}\sum_{n=0}^{N-1}[\underbrace{\theta_1\left(\Delta x-\Delta x_{des}^{EV}\right)^2}_{\text{driving safety}}+\underbrace{\theta_2\left(v_{EV}-v_{des}^{EV}\right)^2}_{\text{travel efficiency}}+\underbrace{\theta_3 a_{EV}^2}_{\text{ride comfort}}] \tag{10}$$

Where $N$ denotes the optimization steps in the optimization horizon $T$ and $n$ denotes the current optimization step; $\Delta x_{des}^{EV}$ represents the desired longitudinal safe distance of EV while $v_{des}^{EV}$ represents the desired speed of EV. The first term in the integral denotes EV's driving safety feature; The second term denotes the mobility feature; The last term mainly specifies ride comfort. $\theta_1,\theta_2,\theta_3$ are weighting factors of each term according to EV users' preference. The behavior model of CV is defined as:

$$J_{CV}=\frac{1}{2}\sum_{n=0}^{N-1}[\underbrace{\beta_1\left(\Delta x_n-\Delta x_{des}^{CV}\right)^2}_{\text{driving safety}}+\underbrace{\beta_2\left(v_{CV,n}-v_{des}^{CV}\right)^2}_{\text{driving efficiency}}+\underbrace{\beta_3 a_{CV,n}^2}_{\text{ride comfort}}$$
$$+\underbrace{\beta_4\left(y_{CV,n}-y_{des}^{CV}\right)^2}_{\text{lane-changing request}}+\underbrace{\beta_5\psi_{CV,n}^2}_{\text{motion smoothness}}\ ] \tag{11a}$$

$$\boldsymbol{\beta}=(\beta_1,\beta_2,\beta_3,\beta_4,\beta_5)^{\mathrm{T}} \tag{11b}$$

where $\Delta x_{des}^{CV}$ represents the desired longitudinal safe distance of CV while $v_{des}^{CV}$ represents the desired speed of CV; $y_{des}^{CV}$ denotes the desired lateral position of CV. The first three terms in equation (11a) driving safety feature, driving efficiency feature, and ride comfort feature of CV. They are longitudinal driving features. The fourth term in equation (11a) denotes CV's lane-change requests. If CV is tended to cut in EV, $y_{des}^{CV}$ denotes the y coordinate of the target lane centerline. The fifth term can reduce the change of yaw angle to ensure CV's motion smoothness. The last two terms are lateral driving features.

## B. CV individual driving style identification

To identify the driving style of CV, a novel algorithm is applied in this research inspired by online inverse optimal control (IOC) [42]. Online IOC is suitable as it can estimate the given parameters through real-time data. This research assumes that the CV follows an MPC behavior model with a predefined cost function, which governs its behavior. The driving style parameters serve as the cost weights in the cost function. Utilizing data including EV states and CV trajectories, this algorithm can recognize CV's preference for each term in the cost function, thus learning the knowledge about CV's individual driving style parameters.

*1) Definition of CV's individual driving style $\boldsymbol{\beta}$*

This research defines $\boldsymbol{\beta}$ as weighting factors of CV behavior model's cost function. As illustrated in (11a) and (11b), the parameters $\beta_1,\beta_2,\beta_3,\beta_4,\beta_5$ determine the individual driving styles of CV as they can reflect CV's preference to different driving features. The equation (11a) can be further transformed to the following form:

$$J_{CV}=\frac{1}{2}\sum_{n=0}^{N-1}\begin{pmatrix}\beta_1\\\beta_2\\\beta_3\end{pmatrix}^{\mathrm{T}}\begin{pmatrix}(\Delta x_n-\Delta x_{des}^{CV})^2\\\left(v_{CV,n}-v_{des}^{CV}\right)^2\\a_{CV,n}^2\end{pmatrix}+\begin{pmatrix}\beta_4\\\beta_5\end{pmatrix}^{\mathrm{T}}\begin{pmatrix}\left(y_{CV,n}-y_{des}^{CV}\right)^2\\\psi_{CV,n}^2\end{pmatrix}$$
$$=\underbrace{\frac{1}{2}\sum_{n=0}^{N-1}\boldsymbol{\beta}_{long}^{\mathrm{T}}\,\boldsymbol{L}_{long,n}\left(\boldsymbol{x}_{long,n},a_{CV,n}\right)}_{\text{longitudinal components}}+\underbrace{\frac{1}{2}\sum_{n=0}^{N-1}\boldsymbol{\beta}_{lat}^{\mathrm{T}}\boldsymbol{L}_{lat,n}\left(\boldsymbol{x}_{lat,n}\right)}_{\text{lateral components}} \tag{12}$$

$$\boldsymbol{x}_{long,n}=(\Delta x_n,v_{EV,n},v_{CV,n})^{\mathrm{T}} \tag{13a}$$

$$\boldsymbol{x}_{lat,n}=(y_{CV,n},\psi_{CV,n})^{\mathrm{T}} \tag{13b}$$

where $\boldsymbol{\beta}_{long}$ represents parameters reflecting longitudinal driving features, including $\beta_1,\beta_2,\beta_3$; $\boldsymbol{\beta}_{lat}$ represents CV's preference to lateral driving features, including $\beta_4,\beta_5$; $\boldsymbol{x}_{long,n}$ and $\boldsymbol{x}_{lat,n}$ contains the longitudinal and lateral components of system state $\boldsymbol{x}$ at step $n$, respectively.

*2) Algorithm formulation*

*Overview.* This paper assumes the CV's observed trajectory is optimal for its unknown cost function weights $\boldsymbol{\beta}$ (also CV's individual driving style) in equation (11a). Using Pontryagin's Minimum Principle, we can derive a necessary condition that this optimal trajectory must satisfy equation (14). We then solve for the $\boldsymbol{\beta}$ that best fits this condition given the observed data. As the lateral and longitudinal movement of vehicles follows different movement modes, this research separately identifies the parameters $\boldsymbol{\beta}_{long}$ and $\boldsymbol{\beta}_{lat}$ using the same algorithm. Due to paper length limitations, this research takes $\boldsymbol{\beta}_{long}$ identification as an example to formulate the algorithm.

***Theorem 1.*** Given the current time step $n$ and the real-time trajectory data of CV, the CV's driving style parameters $\boldsymbol{\beta}_{long}$ should satisfy the linear equation as follows:

$$\boldsymbol{\mathcal{L}}_n\boldsymbol{\mathcal{K}}_n\boldsymbol{\gamma}_0=0 \tag{14}$$

where,

$$\boldsymbol{\mathcal{L}}_n=\begin{pmatrix}\nabla_{a_{CV,n}}\boldsymbol{L}_{long,n} & \boldsymbol{C}_{n+1}^{\mathrm{T}}\end{pmatrix} \tag{15a}$$

$$\boldsymbol{\mathcal{K}}_n=\boldsymbol{K}_n\boldsymbol{K}_{n-1}\boldsymbol{K}_{n-2}\dots\boldsymbol{K}_0 \tag{15b}$$

$$\boldsymbol{K}_n=\begin{pmatrix}\boldsymbol{I} & \boldsymbol{0}\\-(\boldsymbol{A}_n^{\mathrm{T}})^{-1}\nabla_{\boldsymbol{x}_{long,n}}\boldsymbol{L}_{long,n} & (\boldsymbol{A}_n^{\mathrm{T}})^{-1}\end{pmatrix} \tag{15c}$$

$$\boldsymbol{L}_{long,n}=\begin{pmatrix}\left(\Delta x_n-\Delta x_{des}^{CV}\right)^2 & \left(v_{CV,n}-v_{des}^{CV}\right)^2 & a_{CV,n}^2\end{pmatrix}^{\mathrm{T}} \tag{15d}$$

$$\boldsymbol{\gamma}_0=(\boldsymbol{\beta}_{long},\boldsymbol{\lambda}_0)^{\mathrm{T}} \tag{15e}$$

$\boldsymbol{\lambda}_0$ is the initial costate of $\boldsymbol{x}_0$.

***Proof.*** Due to considerations of the complexity of the methodology, the proof is included in Appendix A. The lateral driving style parameters $\boldsymbol{\beta}_{lat}$ follow the same derivation process as $\boldsymbol{\beta}_{long}$.

### 3) Solution method and online implementation

This subsubsection will introduce the solution method of $\boldsymbol{\beta}$ and the online implementation scheme. The solution method and the online implementation scheme is suitable for both $\boldsymbol{\beta}_{long}$ and $\boldsymbol{\beta}_{lat}$. This research takes $\boldsymbol{\beta}_{long}$ as an example in follows.

Due to the noise disturbance or the modeling error, it is difficult to find a set of parameters enabling the collected trajectory data to strictly satisfy equation (14) (Theorem 1). Based on this fact, the equation (14) could be transformed to the following iterative least-squares optimization problem:

$$\begin{aligned} &\min_{\boldsymbol{\gamma}_0} \sum_{n} ||\boldsymbol{\mathcal{L}}_n \boldsymbol{\mathcal{K}}_n \boldsymbol{\gamma}_0||^2 \\ &= \min_{\boldsymbol{\gamma}_0} \boldsymbol{\gamma}_0 \left( \sum_{n} (\boldsymbol{\mathcal{L}}_n \boldsymbol{\mathcal{K}}_n)^{\mathrm{T}} (\boldsymbol{\mathcal{L}}_n \boldsymbol{\mathcal{K}}_n) \right) \boldsymbol{\gamma}_0{}^{\mathrm{T}} \\ &= \min_{\boldsymbol{\gamma}_0} \boldsymbol{\gamma}_0 \boldsymbol{\mathcal{P}}_n \boldsymbol{\gamma}_0{}^{\mathrm{T}} \end{aligned} \tag{16}$$

$$s.t.\ \boldsymbol{e}\boldsymbol{\gamma}_0 = \beta_1 = \imath \tag{17}$$

where $\boldsymbol{\mathcal{P}}_n$ is a positive semidefinite matrix; $\boldsymbol{e} = (1,0,0,\dots,0)$; $\imath$ is a scaling factor to let the first component of $\boldsymbol{\gamma}_0$, which is $\beta_1$,equal to $\imath$. Based on the Lagrange multiplier method, the following equation could be obtained:

$$2\boldsymbol{\mathcal{P}}_n \boldsymbol{\gamma}_0 + p\boldsymbol{e}^{\mathrm{T}} = \boldsymbol{0} \tag{18}$$

where $p$ is the Lagrange multiplier. $\boldsymbol{\beta}_{long}$ can be derived by solving equation (18). we do not restrict non-negativity in $\boldsymbol{\gamma}_0$ except for $\beta_1$. [42] has succinctly proven that as more data is input, $\boldsymbol{\gamma}_0$ will definitely converges to a finite value. The algorithm for identifying individual driving styles is summarized as an online implementation scheme illustrated in Figure 4. This algorithm is implemented in an iterative way. At each step, EV collects the real-time trajectory data of CV online and calculates $\boldsymbol{\beta}^*_{long}$. Upon $\boldsymbol{\beta}^*_{long}$ has converged, the algorithm outputs the optimal parameters and terminates.

### 4) Numerical simulation on real-time identification

The effectiveness of the CV individual driving style identification algorithm is validated through a numerical simulation. This research defines two parameters set $\boldsymbol{\beta}$ of CV's cost functions to simulate two types of driving styles. To be detailed, $\boldsymbol{\beta} = (1, 5, 0.1, 1, 0.1)^{\mathrm{T}}$ simulates the aggressive CV drivers, which means the drivers prefer efficiency ($\beta_2$) more and overlook the comfort ($\beta_3$) and motion smoothness ( $\beta_5$ ). By contrast, $\boldsymbol{\beta} = (1, 1, 10, 1, 5)^{\mathrm{T}}$ simulates the conservative drivers. $\beta_1$ and $\beta_4$ are set to 1 because they serve as the scalar in the longitudinal and lateral dimension respectively. Through solving equation (11a), the trajectory data of CV is obtained and fed into the proposed individual driving style identification algorithm. Figure 5 presents the aggressive driving style identification results. The parameters converge in 1.1s and achieve acceptable accuracy. The identified $\beta_2$ finally converges to 5.46 after 6 iteration steps (0.6s). The identified $\beta_3$ finally converges to 0.104 after 11 iteration steps (1.1s). The identified $\beta_5$ finally converges to 0.13 after 11 iteration steps (1.1s). Figure 6 presents the conservative driving style identification results. The parameters converge in 1.3s and also achieve acceptable accuracy. The identified $\beta_2$ finally converges to 1.14 after 11 iteration steps (1.1s). The identified $\beta_3$ finally converges to 10.21 after 5 iteration steps (0.5s). The identified $\beta_5$ finally converges to 5.57 after 13 iteration steps (1.3s).

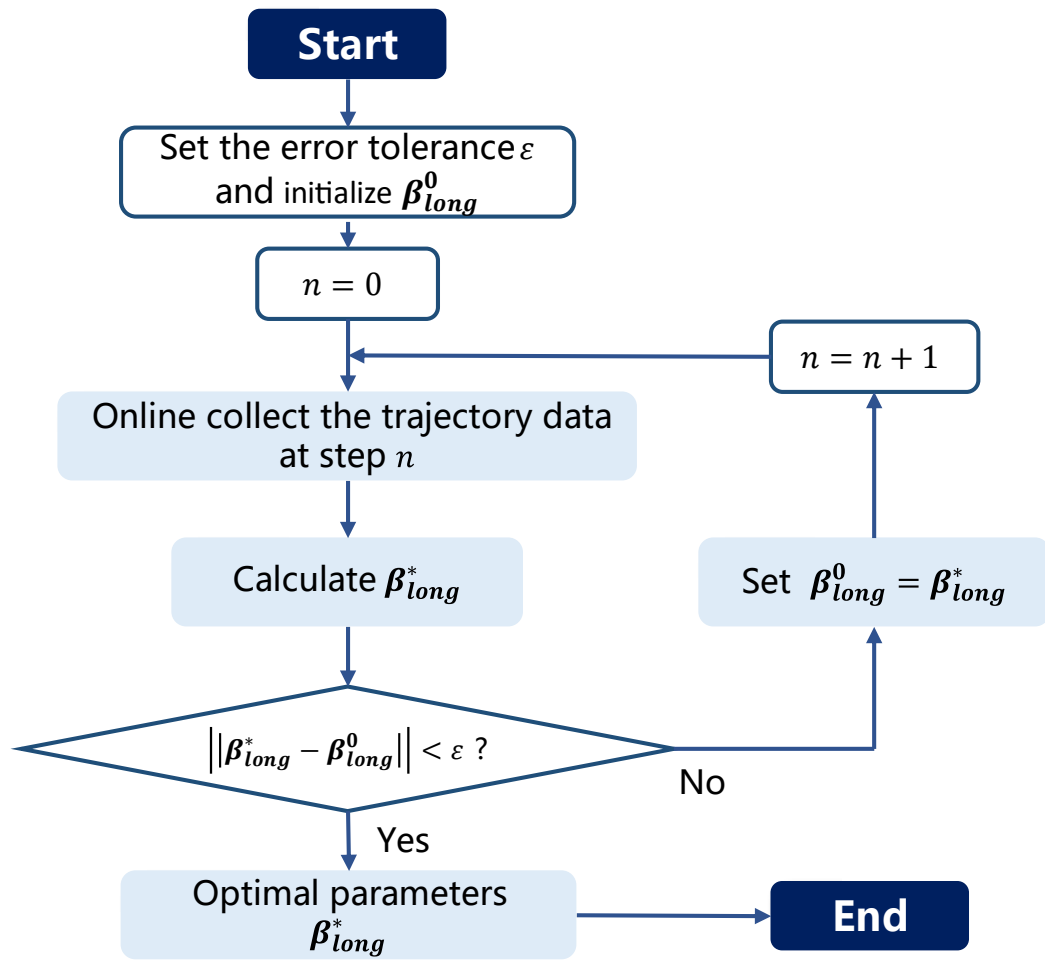

Figure 4 Online implementation scheme of the algorithm. At each step, EV collects the real-time trajectory data of CV online and calculates $\boldsymbol{\beta}^*_{long}$. Upon $\boldsymbol{\beta}^*_{long}$ has converged, the algorithm outputs the optimal parameters and terminates.

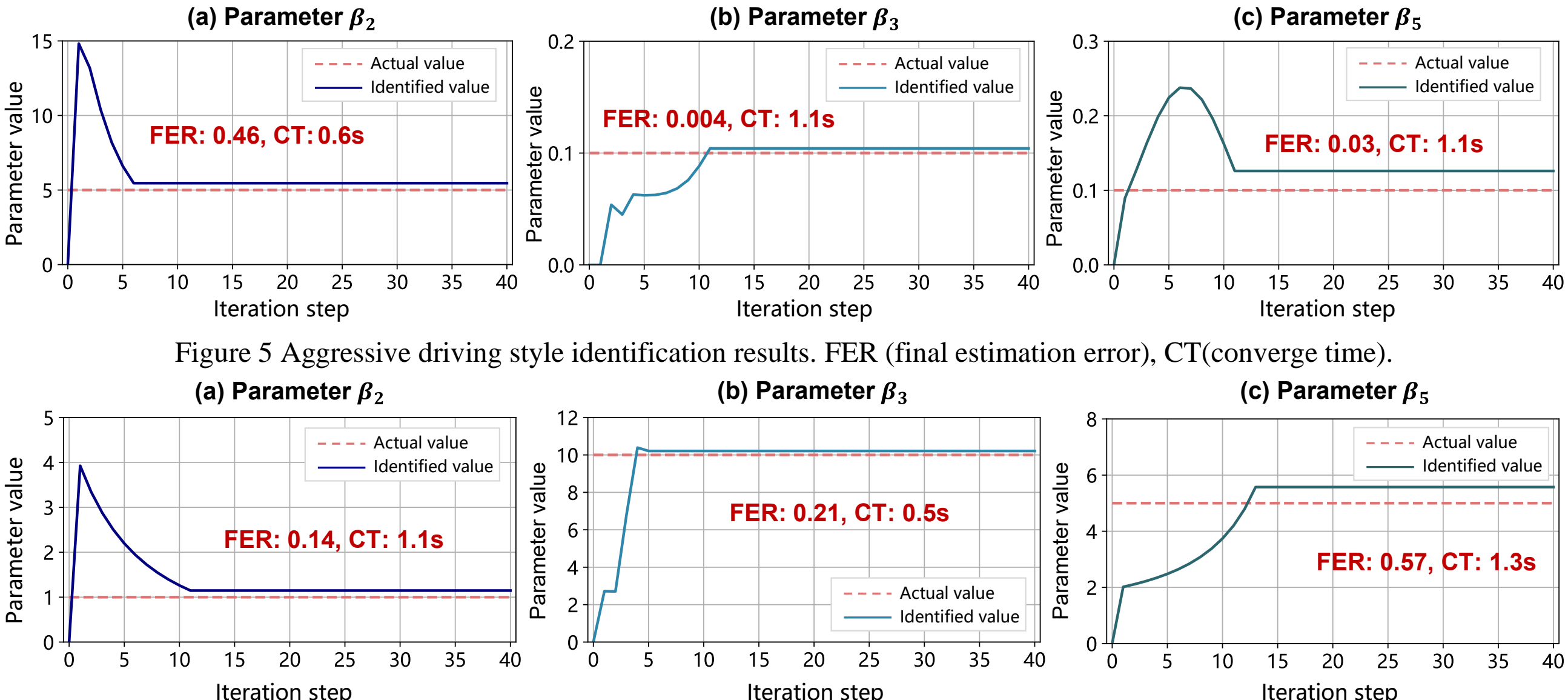

Figure 5 Aggressive driving style identification results. FER (final estimation error), CT(converge time).

Figure 6 Conservative driving style identification results. FER (final estimation error), CT(converge time).

### C. Game-theoretic-based motion planning framework

*Overview.* This research designed a Game-theoretic-based Model Predictive Control (GMPC) planner for the right-of-way protection of EV. Stackelberg competition is adopted as the game rule [44] as it can reflect the leader–follower dynamics between the EV and the CV, making it a natural fit for modeling such asymmetric interactions. In Stackelberg competition, the leader takes action first, and the follower reacts to the leader's action based on observations. For this research, EV could be accounted as a leader and proactively compete with the adjacent CV for right-of-way protection; as a follower, CV would react to EV's action.

*Strategy.* Figure 7 illustrates the implementation details of the Stackelberg-competition-based planner framework. In the CV reaction estimation module, this approach assumes that CV follows a behavioral model, which is formulated as a MPC problem. The reaction function can be obtained by solving the MPC problem by dynamic programming. In GMPC planner module, this approach utilizes the optimal reaction function of CV to build the EV-CV interaction system dynamics constraint and optimizes reference command for EV. After obtaining the reference command from the cloud, EV, as the leader, takes action first (action ① in Figure 7 (III)). Then, CV, as the follower, conducts a reaction (action ② in Figure 7 (IV)), completing the interactive cycle between the leader and the follower in the Stackelberg game context. Finally, the whole process runs to the next time step to update $\boldsymbol{\beta}$.

#### 1) CV reaction estimation

*Overview.* Based on the Stackelberg competition

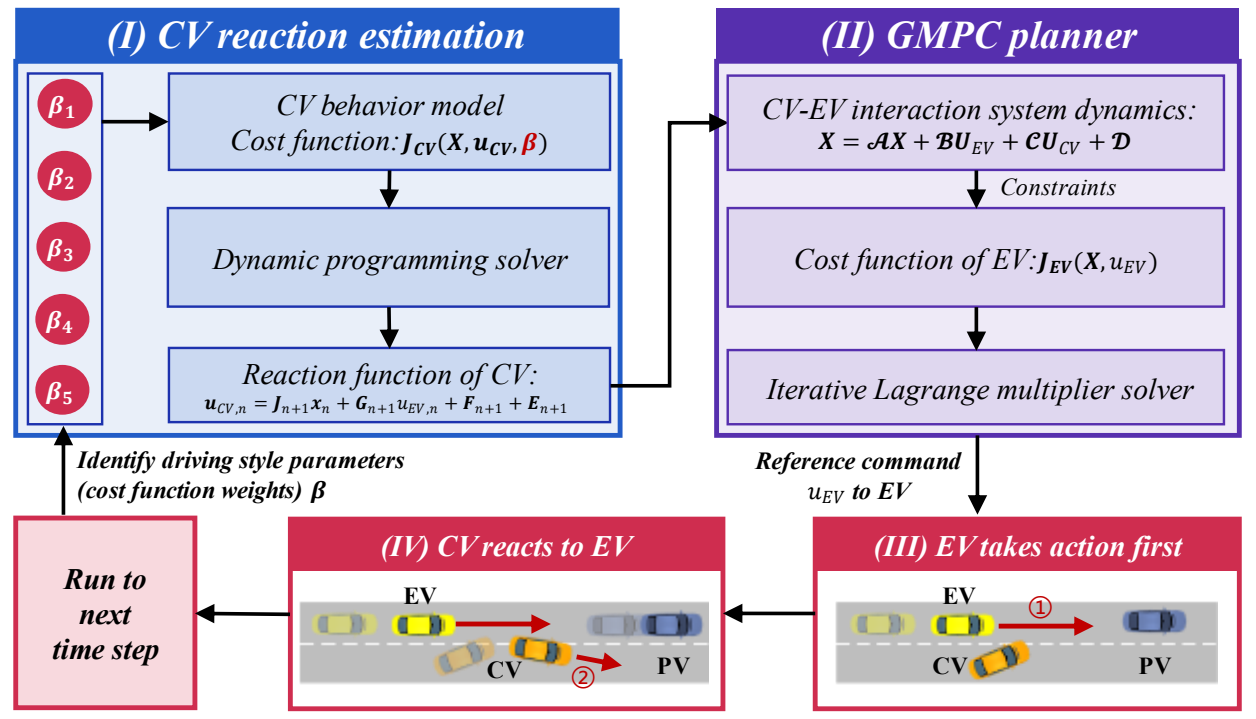

Figure 7 Illustration of Stackelberg-competition-based planner implementation framework. We assume that EV takes action first and CV conducts a reaction, completing the interactive cycle between the leader and the follower in the Stackelberg game context.

framework, CV, as the follower, moves according to its observation of EV. Hence, the reaction $\boldsymbol{u}_{CV,n}$ of CV can be estimated according to system state vector $\boldsymbol{x}_n$, EV's reference command $u_{EV,n}$, and the identified individual driving style parameters $\boldsymbol{\beta}$ as shown in equation (19). To derive $\boldsymbol{u}_{CV,n}$, we solve the behavior model equation (11a) of CV through the dynamic programming.

***Theorem 2.*** Assuming the reaction of CV follows an optimal control scheme, the optimal reaction function of CV in responding to EV's action is:

$$\boldsymbol{u}_{CV,n} = \boldsymbol{J}_{n+1}\boldsymbol{x}_n + \boldsymbol{G}_{n+1}u_{EV,n} + \boldsymbol{F}_{n+1} + \boldsymbol{E}_{n+1} (n = 0, \dots, N-1) \quad (19)$$

where,

$$\boldsymbol{J}_{n+1} = \boldsymbol{H}_{n+1}\boldsymbol{M}_{n+1}\boldsymbol{A}_n \quad (20a)$$

$$\boldsymbol{G}_{n+1} = \boldsymbol{H}_{n+1}\boldsymbol{M}_{n+1}\boldsymbol{B}_n \quad (20b)$$

$$\boldsymbol{F}_{n+1} = \boldsymbol{H}_{n+1}\boldsymbol{M}_{n+1}\boldsymbol{x}_{des}^{CV} \quad (20c)$$

$$\boldsymbol{E}_{n+1} = \boldsymbol{H}_{n+1}\boldsymbol{O}_{n+1} \quad (20d)$$

$$\boldsymbol{H}_{n+1} = -(\boldsymbol{C}_n^{\mathrm{T}}\boldsymbol{M}_{n+1}\boldsymbol{C}_n + \boldsymbol{\beta}_d)^{-1}\boldsymbol{C}_n^{\mathrm{T}} \quad (20e)$$

$$\boldsymbol{x}_{des}^{CV} = (\Delta x_{des}^{CV}, 0, v_{des}^{CV}, y_{des}^{CV}, \psi_{des}^{CV})^{\mathrm{T}} \quad (20f)$$

$\boldsymbol{M}_n$, $\boldsymbol{O}_n$ could be obtained according to $\boldsymbol{M}_N$, $\boldsymbol{O}_N$ and the following recursions in reverse order:

$$\boldsymbol{O}_N = (0,0,0,0,0)^{\mathrm{T}} \quad (21a)$$

$$\boldsymbol{M}_N = diag(0,0,0,0,0) \quad (21b)$$

$$\boldsymbol{O}_n = \boldsymbol{A}_n^{\mathrm{T}}(\boldsymbol{I} + \boldsymbol{C}_n\boldsymbol{H}_{n+1}\boldsymbol{M}_{n+1})^{\mathrm{T}}(\boldsymbol{O}_{n+1} + \boldsymbol{M}_{n+1}\boldsymbol{B}_n\boldsymbol{u}_{EV,n}) \quad (21c)$$

$$\boldsymbol{M}_n = \boldsymbol{A}_n^{\mathrm{T}}\boldsymbol{M}_{n+1}\boldsymbol{A}_n + \boldsymbol{Q}_{CV} - \boldsymbol{A}_n^{\mathrm{T}}\boldsymbol{M}_{n+1}\boldsymbol{C}_n\boldsymbol{H}_{n+1}\boldsymbol{M}_{n+1}\boldsymbol{A}_n \quad (21d)$$

$$\boldsymbol{Q}_{CV} = diag(\beta_1, 0, \beta_2, \beta_4, \beta_5) \quad (21e)$$

$$\boldsymbol{\beta}_d = diag(\beta_3, 0) \quad (21f)$$

This linear reaction function in equation (19) explicitly shows how the CV's motions (acceleration, steering) depends on the current state $\boldsymbol{x}_n$, the EV's motion $u_{EV,n}$, and its own driving style $\boldsymbol{\beta}$.

***Proof.*** Due to considerations of the complexity of the methodology, the proof is included in Appendix B.

#### 2) GMPC planner design

*Overview.* The GMPC planner is designed for EV to protect right-of-way. The reward function $\mathcal{R}$ in POMDP framework is defined as EV's cost function (10). The planner is solved in a receding horizon fashion via MPC. According to Stackelberg game rule, CV's reaction formulated in equation (19) is considered in the EV-CV interaction system dynamics constraint.

*Cost function formulation:* the cost function converts equation (10) into the quadratic programming form:

$$J_{EV} = \frac{1}{2}(\boldsymbol{X}^{\mathrm{T}} \quad \boldsymbol{U}_{EV}^{\mathrm{T}})\boldsymbol{\mathcal{M}}\begin{pmatrix}\boldsymbol{X}\\ \boldsymbol{U}_{EV}\end{pmatrix} + \boldsymbol{q}^{\mathrm{T}}\begin{pmatrix}\boldsymbol{X}\\ \boldsymbol{U}_{EV}\end{pmatrix} \quad (22)$$

where:

$$\boldsymbol{X} = (\boldsymbol{x}_0, \boldsymbol{x}_1, \dots, \boldsymbol{x}_N)^{\mathrm{T}} \quad (23a)$$

$$\boldsymbol{U}_{EV} = (u_{EV,0}, u_{EV,1}, \dots, u_{EV,N-1})^{\mathrm{T}} \quad (23b)$$

$$\boldsymbol{\mathcal{M}} = diag\,(\underbrace{\boldsymbol{Q}_{EV}, \dots, \boldsymbol{Q}_{EV}}_{N}, \boldsymbol{0}_{5\times5}, \underbrace{\theta_3, \dots, \theta_3}_{N}) \quad (23c)$$

$$\boldsymbol{Q}_{EV} = diag(\theta_1, \theta_2, 0,0,0) \tag{23d}$$

$$\boldsymbol{q} = (\underbrace{-\boldsymbol{Q}_{EV}\boldsymbol{x}_{des}^{EV}, \dots, -\boldsymbol{Q}_{EV}\boldsymbol{x}_{des}^{EV}}_{N}, 0, \dots, 0)^{\mathrm{T}} \tag{23e}$$

$$\boldsymbol{x}_{des}^{EV} = (\Delta x_{des}^{EV}, v_{des}^{EV}, 0,0,0)^{\mathrm{T}} \tag{23f}$$

*EV-CV interaction system dynamics constraint:* the EV-CV interaction system dynamics constraint of EV can be transformed from equation (8):

$$\boldsymbol{X} = \boldsymbol{\mathcal{A}}\boldsymbol{X} + \boldsymbol{\mathcal{B}}\boldsymbol{U}_{EV} + \boldsymbol{\mathcal{C}}\boldsymbol{U}_{CV} + \boldsymbol{\mathcal{D}} \tag{24}$$

where,

$$\boldsymbol{\mathcal{A}} = \begin{bmatrix} \boldsymbol{0} & \boldsymbol{0} & \cdots & \boldsymbol{0} & \boldsymbol{0} \\ \boldsymbol{A}_0 & \boldsymbol{0} & \cdots & \boldsymbol{0} & \boldsymbol{0} \\ \boldsymbol{0} & \ddots & \ddots & \vdots & \vdots \\ \vdots & \boldsymbol{0} & \ddots & \boldsymbol{0} & \vdots \\ \boldsymbol{0} & \boldsymbol{0} & \cdots & \boldsymbol{A}_{N-1} & \boldsymbol{0} \end{bmatrix} \tag{25a}$$

$$\boldsymbol{\mathcal{B}} = \begin{bmatrix} \boldsymbol{0} & \boldsymbol{0} & \cdots & \boldsymbol{0} \\ \boldsymbol{B}_0 & \boldsymbol{0} & \cdots & \boldsymbol{0} \\ \boldsymbol{0} & \ddots & \ddots & \vdots \\ \vdots & \boldsymbol{0} & \ddots & \boldsymbol{0} \\ \boldsymbol{0} & \boldsymbol{0} & \cdots & \boldsymbol{B}_{N-1} \end{bmatrix} \tag{25b}$$

$$\boldsymbol{\mathcal{C}} = \begin{bmatrix} \boldsymbol{0} & \boldsymbol{0} & \cdots & \boldsymbol{0} \\ \boldsymbol{C}_0 & \boldsymbol{0} & \cdots & \boldsymbol{0} \\ \boldsymbol{0} & \ddots & \ddots & \vdots \\ \vdots & \boldsymbol{0} & \ddots & \boldsymbol{0} \\ \boldsymbol{0} & \boldsymbol{0} & \cdots & \boldsymbol{C}_{N-1} \end{bmatrix} \tag{25c}$$

$$\boldsymbol{\mathcal{D}} = (\boldsymbol{x}_0, 0, \dots, 0)^{\mathrm{T}} \tag{25d}$$

$$\boldsymbol{U}_{CV} = (\boldsymbol{u}_{CV,0}, \boldsymbol{u}_{CV,1}, \dots, \boldsymbol{u}_{CV,N-1})^{\mathrm{T}} \tag{25e}$$

substituting equation (19) to equation (25e), $\boldsymbol{U}_{CV}$ can be represented as:

$$\boldsymbol{U}_{CV} = \boldsymbol{\mathcal{J}}\boldsymbol{X} + (\boldsymbol{\mathcal{G}} + \boldsymbol{\mathcal{E}})\boldsymbol{U}_{EV} + \boldsymbol{\mathcal{F}} \tag{26}$$

where,

$$\boldsymbol{\mathcal{J}} = \begin{bmatrix} \boldsymbol{0} & \boldsymbol{0} & \cdots & \boldsymbol{0} & \boldsymbol{0} \\ \boldsymbol{J}_1 & \boldsymbol{0} & \cdots & \boldsymbol{0} & \boldsymbol{0} \\ \boldsymbol{0} & \ddots & \ddots & \vdots & \vdots \\ \vdots & \boldsymbol{0} & \ddots & \boldsymbol{0} & \vdots \\ \boldsymbol{0} & \boldsymbol{0} & \cdots & \boldsymbol{J}_N & \boldsymbol{0} \end{bmatrix} \tag{27a}$$

$$\boldsymbol{\mathcal{G}} = diag(\boldsymbol{G}_1, \boldsymbol{G}_2, \dots, \boldsymbol{G}_N) \tag{27b}$$

$$\boldsymbol{\mathcal{F}} = diag(\boldsymbol{F}_1, \boldsymbol{F}_2, \dots, \boldsymbol{F}_N) \tag{27c}$$

$$\boldsymbol{\mathcal{E}} = \begin{bmatrix} \boldsymbol{0}_{(N-1)\times 2} & \widetilde{\boldsymbol{\mathcal{E}}} \\ \boldsymbol{0}_{1\times 2} & \boldsymbol{0}_{1\times 2(N-1)} \end{bmatrix} \tag{27d}$$

$$\widetilde{\boldsymbol{\mathcal{E}}}_{nj} = \begin{cases} \boldsymbol{H}_n \prod_{i=n+1}^{j} \boldsymbol{S}_i \boldsymbol{T}_{j+1} & , n < j \\ \boldsymbol{H}_n \boldsymbol{T}_{n+1} & , n = j \\ \boldsymbol{0}_{1\times 2} & , n < j \end{cases} \tag{27e}$$

$$\boldsymbol{S}_{n+1} = \boldsymbol{A}_n^{\mathrm{T}}(\boldsymbol{I} - \boldsymbol{C}_n(\boldsymbol{C}_n^{\mathrm{T}}\boldsymbol{M}_{n+1}\boldsymbol{C}_n + \boldsymbol{\beta}_d)^{-1}\boldsymbol{C}_n^{\mathrm{T}}\boldsymbol{M}_{n+1})^{\mathrm{T}} \tag{27f}$$

$$\boldsymbol{T}_{n+1} = \boldsymbol{S}_{n+1}\boldsymbol{M}_{n+1}\boldsymbol{B}_n \tag{27g}$$

then, substitute equation (26) to equation (24), the EV-CV interaction system dynamics constraint of EV can be expressed as:

$$\begin{pmatrix} \boldsymbol{I} - (\boldsymbol{\mathcal{A}} + \boldsymbol{\mathcal{C}}\boldsymbol{\mathcal{J}}) & -\boldsymbol{\mathcal{C}}(\boldsymbol{\mathcal{G}} + \boldsymbol{\mathcal{E}}) - \boldsymbol{\mathcal{B}} \end{pmatrix} \begin{pmatrix} \boldsymbol{X} \\ \boldsymbol{U}_{EV} \end{pmatrix} = \boldsymbol{\mathcal{C}}\boldsymbol{\mathcal{F}} + \boldsymbol{\mathcal{D}} \tag{28}$$

To this end, by denoting $\boldsymbol{U}_{CV}$ by $\boldsymbol{U}_{EV}$, the EV-CV interaction system dynamics constraint is only about $\boldsymbol{U}_{EV}$.

*Collision avoidance constraint:* To mitigate potential risks raised from cut-in prevention, a chance constraint is added for collision avoidance using the big-M method for computational efficiency. The M parameter is 20, which is sufficient to cover the reachable relative longitudinal and lateral ranges. This value is also chosen to avoid the computational drawbacks caused by an excessively large M. To enhance readability, we use pseudocode to illustrate how these constraints are added as illustrated in TABLE 1.

TABLE 1 Collision avoidance constraint adding process

| | **Adding the collision avoidance constraints** |
|---|---|
| 1 | **Initialize:** a large enough positive number M; The centerline lateral position $y_{EV}$ of the lane EV occupied; longitudinal safety distance reference $\Delta x_{ref}$; lateral safety distance reference $\Delta y_{ref}$; planning horizon $N$ |
| 2 | **for** $n \leftarrow 1$ to $N$ **do:** |
| 3 | Initialize a binary variable $c_{EV,n} \in \{0,1\}$ |
| | Adding constraints to the planner: |
| 4 | $\lvert\Delta x_n\rvert \geq \Delta x_{ref} - \mathrm{M} \cdot c_{EV,n}$<br>$\lvert y_{EV} - y_{CV,n}\rvert \geq \Delta y_{ref} - \mathrm{M} \cdot (1 - c_{EV,n})$ |
| 5 | **end for** |

*Other constraints:* This research also considers other constraints such as speed range, action range, and initial conditions.

*Solution method:* according to the quadratic cost function equation (22), the linear EV-CV interaction system dynamics constraint equation (28), and other constraints. The optimization problem of the GMPC planner is easy to be solved by current solvers [45][46].

## IV. Evaluation

The proposed vehicle-cloud cooperative anti-bullying adaptive cruise control (AACC) approach is assessed and evaluated based on the following aspects: i) "road-bullying" (cut-ins) prevention capability; ii) different CV driving styles adaptation validation; iii) driving comfort, safety, efficiency, and flexibility validation; iv) computational efficiency validation. The evaluation contains two parts: "road bullying" prevention function validation and traffic simulation.

### *A. Preparation*

The preparation for the performance evaluation is detailed introduced, including driving style definition, control types for comparison, and parameter settings.

#### *1) Driving styles definition*

In order to demonstrate that the proposed approach is adaptive to different driving styles, CVs with different driving styles are considered. This research categorized the driving styles of CVs into two categories: conservative and aggressive. This research simulates the longitudinal and lateral behavior of different driving styles using the Intelligent Driver Model (IDM) [47] and Minimizing Overall Braking Induced by Lane-change (MOBIL) model [48]. By tuning the parameters of IDM and MOBIL, the driving styles are defined as follows:

- *Conservative:* drivers with a conservative driving style pay more attention to safety and comfort performances rather than efficiency. For IDM, the tuned parameters are: maximum acceleration $a_{max}^{con} = 1m/s^2$ , comfortable braking deceleration $b_{com}^{con} = 2m/s^2$ , desired time gap $h_{con} = 2.5s$ , desired speed $v_{des}^{CV} = 18m/s$ ; and the parameters for MOBIL are: politeness factor $p_{con} = 0.2$, and lane-changing decision threshold at $a_{th}^{con} = 0.4m/s^2$.
- *Aggressive:* driving requirements of the aggressive drivers are the opposite of conservative drivers. The IDM parameters are: $a_{max}^{agg} = 2.5m/s^2$ , $b_{com}^{agg} = 3m/s^2$ , $h_{agg} = 0.8s$ , and $v_{des}^{CV} = 25m/s$ ; and for MOBIL: $p_{agg} = 0.05$ and $a_{th}^{agg} = 0.2m/s^2$.
- *Neutral:* driving requirements of the neutral drivers are built on the intermediate states. The IDM parameters are: $a_{max}^{neu} = 1.8m/s^2$ , $b_{com}^{neu} = 2.5m/s^2$ , $h_{neu} = 1.5s$ , and $v_{des}^{CV} = 21m/s$; and for MOBIL: $p_{neu} = 0.12$ and $a_{th}^{neu} = 0.3m/s^2$.

*2) Control types for comparison*

In this experiment, three control types are considered:

- *Proposed AACC approach*: in this case, EV is able to online identify the individual driving style of its CV, then utilizing GMPC planner to proactively protect the right-of-way.
- *Baseline traditional ACC approach*: it is a traditional car-following ACC approach and is unable to proactively protect the right-of-way. It always chooses to decelerate and yield when being cut in. The algorithm is borrowed from [49] .
- *Baseline NE-MPC approach*: it is a state-of-the-art approach proposed by Meng et al, [50]. It depends on a Nash-Equilibrium-based game-theoretic MPC framework to handle driver cut-ins. It cannot identify drivers' individual driving styles.

*3) Parameter settings*

The following settings are adopted for the experiment evaluation:

- The length of the freeway is 3 km.
- Lane width is 3.5m.
- Time step is 0.1 s.
- Optimization horizon is 1 s.
- The speed limit is 25 m/s.
- The acceleration range is [-3.5, 4] $m/s^2$.
- The desired speed of EV and CV is 18m/s.
- The desired longitudinal safe distance of CV is 25 m.
- The desired longitudinal safe distance of EV is 0 m when identifying CV as conservative and 25 m when aggressive.
- Weighting factors $\theta_1, \theta_2, \theta_3$ is 10, 10, and 1 respectively.
- $l_r$ and $l_f$ are both 2 m.

*B. "Road bullying" prevention function validation*

*1) Experiment design*

*Testbed*: The function validation platform is a Prescan and Matlab/Simulink incorporated simulation platform. Figure 8 explicitly depicts the structure of the simulation platform. The autonomous driving simulation software Prescan is used to generate the scenario and control EV. The proposed approach is coded with Matlab/Simulink, and served as a motion planning module in Prescan. Prescan is a physics-based simulation platform that replicates realistic vehicle responses, sensor interactions, and environmental dynamics in a controlled virtual environment. A freeway with two lanes is adopted as the testbed in which EV competes with CV for right-of-way protection.

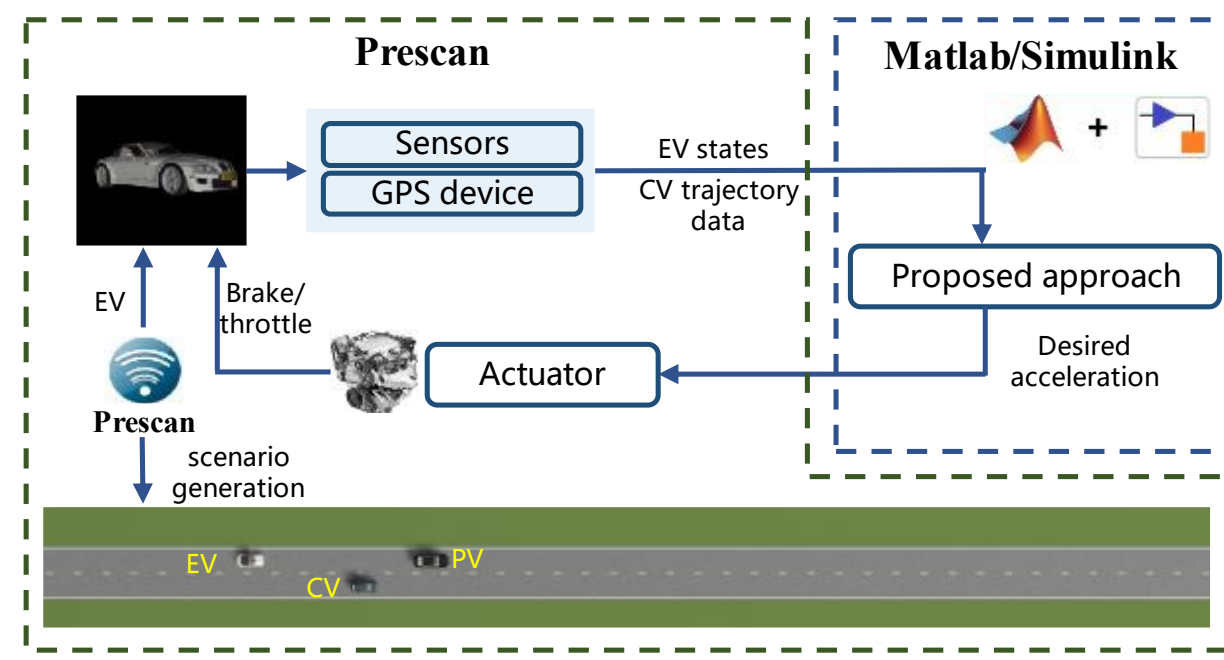

Figure 8 The structure of the simulation platform.

*Sensitive analysis*: Sensitivity analysis considers two factors: driving styles and the initial longitudinal distance between EV and CV. Conservative and aggressive driving styles of CVs are all tested. The initial longitudinal distance between EV and CV contains 10m, 20m, and 30m. It is used to measure the capability of proactively right-of-way protection.

*Metrics*: The following metrics are utilized for performance evaluation:

- Speed, longitudinal position, and sampled trajectories of EV and CV are used to validate right-of-way protection capability and the driving style adaptation ability.
- The Time-integrated Time Headway (TTH) of EV is adopted to quantitively verify the safety improvement of the proposed approach. It is defined as the integration of time headway values of EV below the time headway safety threshold for a period of time. The lower TTH value means the safer performance.
- Coefficient of variation of EV's speed is utilized for evaluating the driving comfort of the proposed approach. The metric is the ratio of speed's standard deviation value to mean value. The lower this metric means the better comfort.

*2) Main results*

The function validation results demonstrate that the proposed approach: i) is capable of preventing "road bullying" cut-ins, thus proactively protecting right-of-way; ii) is adaptive to different driving styles of CV; iii) has benefits in safety performance by relieving the aggravation of potential risks; iv) improves comfort performance by 20.8% in average when CV with conservative driving styles. Across multiple simulation runs, the improvements are statistically significant with $p < 0.05$.

*Right-of-way protection capability*: The proposed approach is proven to have right-of-way protection capability as shown in Figure 9, Figure 10, and Figure 11. Figure 11 presents the results of the baseline traditional ACC. It can be observed that this approach consistently yields to CV, with EV's longitudinal position remaining behind CV. Upon each cut-in event, this approach reduces its speed to adapt to CV's motion and accommodate the cut-in. In contrast, the proposed approach and the baseline NE-MPC approach have right-of-way protection capability. As shown in Figure 9 and Figure 10, both approaches can compete for avoiding being cut in when CV with a conservative driving style or neutral driving style, thus protecting right-of-way. Meanwhile, both approaches can yield to aggressive CV drivers for safety considerations. However, there is an interesting finding when comparing Figure 9 (a) with Figure 10 (a). We found that the proposed approach decelerates at first while the baseline NE-MPC approach accelerates. The zoomed-in region shows that the proposed approach can maintain a larger distance than the baseline NE-MPC approach. This indicates that the proposed approach maintains better right-of-way capability by proactively and predictively avoiding potential conflict. Moreover, three approaches display similar maneuvers when facing conservative and neutral driving styles.

*Different driving styles adaptation ability*: The proposed approach can adapt to different CV driving styles. As shown in Figure 9 (b) and (c), when CV with a conservative driving style, the proposed approach proactively protects the right-of-way of EV by leaving no room for CV to cut in. To be detailed, EV decisively accelerates to overtake and CV reacts to EV's accelerating behavior by decelerating and yielding. Figure 9 (a) presents the validation results when CV with an aggressive style. In this scenario, the proposed approach lowers the safety risk by yielding to CV. To be detailed, EV decelerates predictively when CV aggressively speeds up and cuts in. It can be seen that EV abandons the overtaking maneuver and decelerates to widen the gap from the aggressive CV. Compared with the baseline NE-MPC approach in Figure 10 (a), the proposed approach can identify the aggressive nature of CV and react proactively to maintain a safe distance from CV. But the baseline NE-MPC approach blindly accelerates and reacts at an urgent distance, which induces potential risks.

*Safety performance quantification*: The safety performance of the proposed approach is quantified by TTH in Figure 12 (a), Figure 12 (b), and Figure 12 (c). When CV is with a conservative driving style, Figure 12 (b) shows that the proposed approach has better safety performance under 10m and 20m cut-in gaps compared with the baseline traditional ACC. The TTH benefits are 79.8% and 62.2% respectively. That's because the proposed approach can predictively overtake CV to avoid smaller time headway values due to being behind CV. However, under 30m cut-in gaps, the proposed approach has worse safety performance than the baseline traditional ACC. That's because the overtaking process is riskier as the cut-in gap is longer. Compared with the baseline NE-MPC approach, the proposed approach has

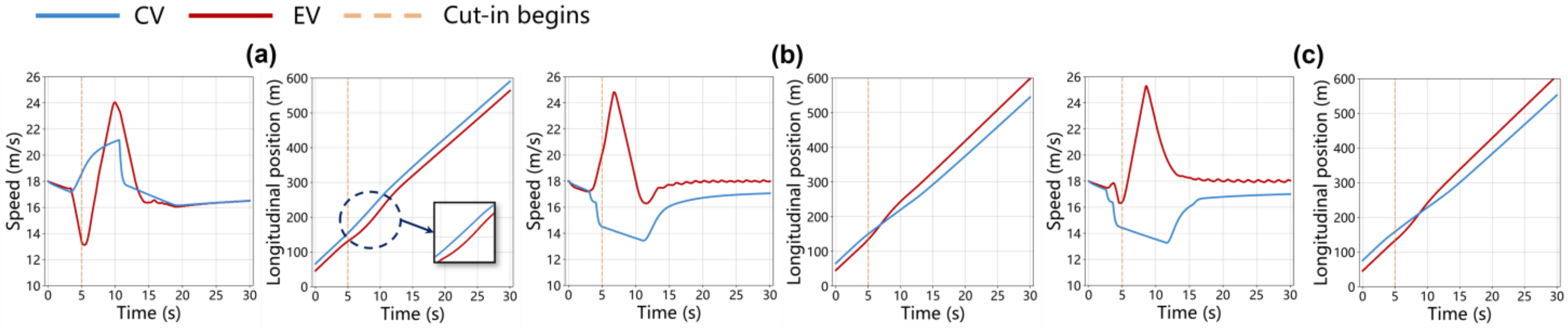

Figure 9 Proposed approach: (a) Trajectories when CV has an aggressive driving style. The proposed approach identifies this and yields early, maintaining a safer distance compared to NE-MPC; In (b) Trajectories when CV has a conservative driving style and (c) Trajectories when CV has a neutral driving style, the proposed approach avoids being cut in and protects its right-of-way.

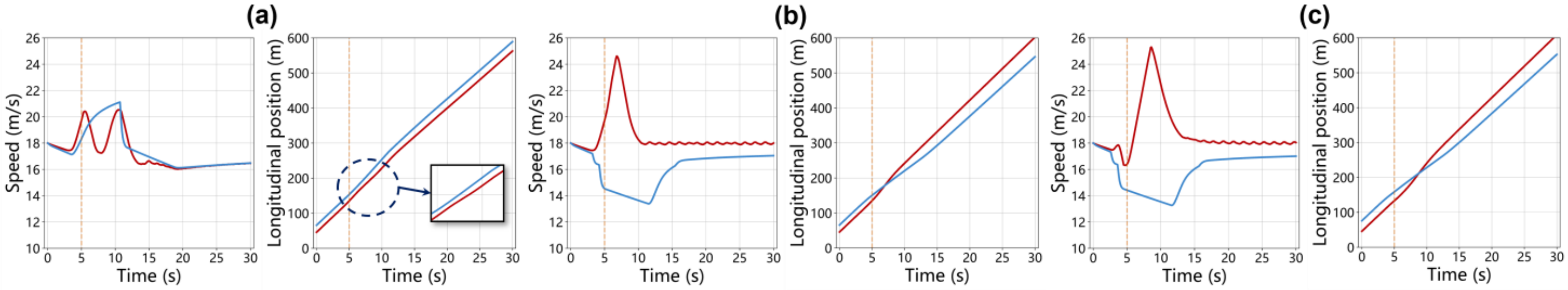

Figure 10 Baseline NE-MPC approach: In (a) Trajectories when CV has an aggressive driving style, this baseline can also yield, but be less safe than the proposed approach; This baseline can also avoid being cut in as shown in (b) Trajectories when CV has a conservative driving style and (c) Trajectories when CV has a neutral driving style.

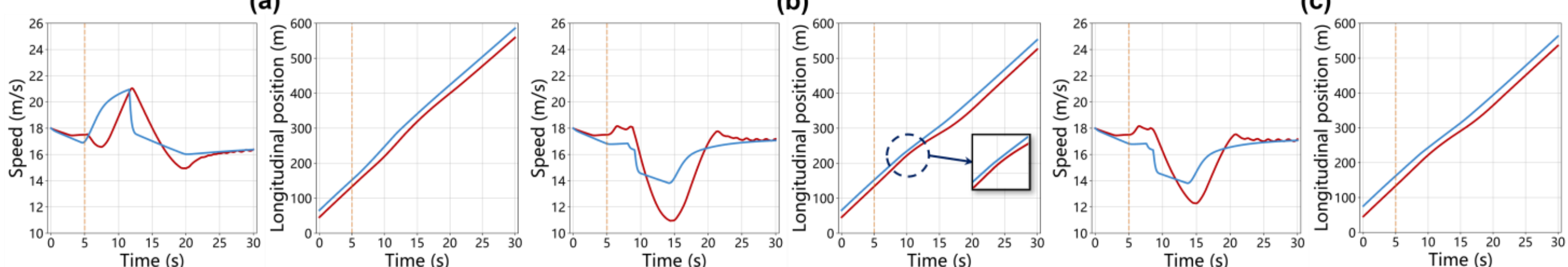

Figure 11 Baseline traditional ACC: (a) Trajectories when CV has an aggressive driving style; (b) Trajectories when CV has a conservative driving style; (c) Trajectories when CV has a neutral driving style. This baseline consistently yields to CV and accommodate the cut-ins.

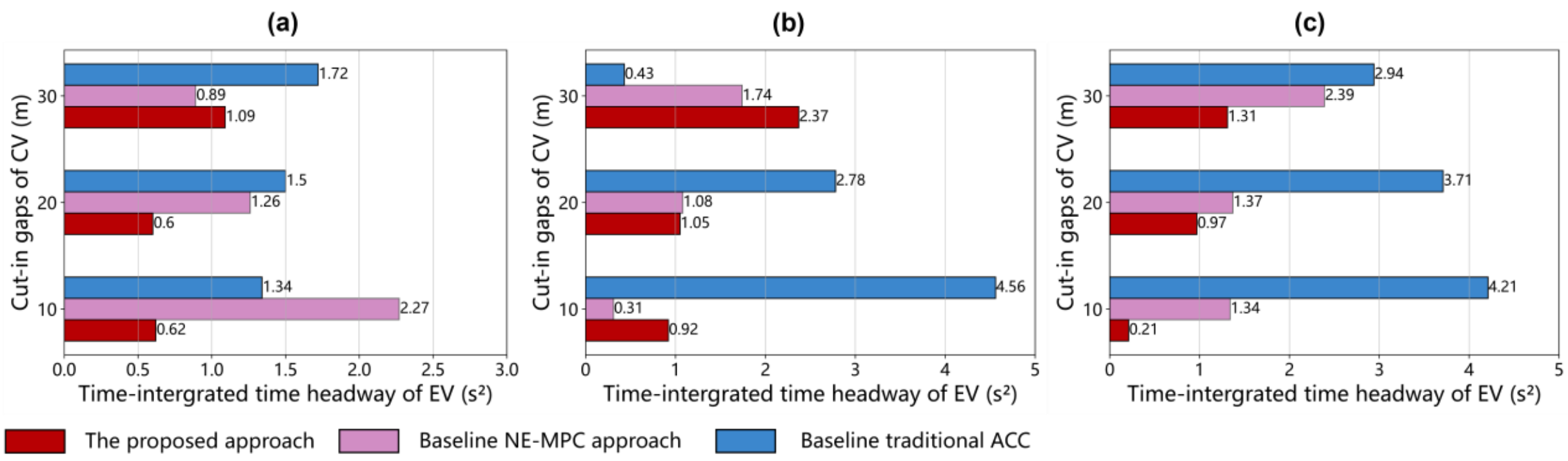

Figure 12 Safety quantification under different cut-in gaps when CV with (a) aggressive driving style; (b) conservative driving style; (c) neutral style.

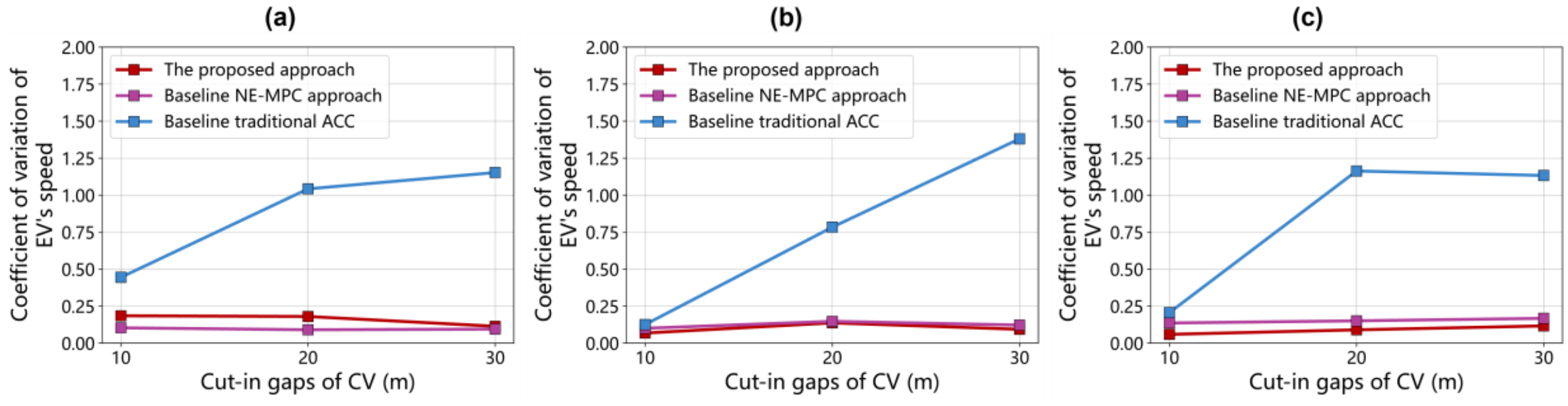

Figure 13 comfort quantification under different cut-in gaps when CV with (a) aggressive driving style; (b) conservative driving style; (c) neutral style.

worse performance in the 10m and 30m cut-in gaps, but not unsafe. That's because the proposed approach is more competitive when identifying the conservative nature of CV.

When CV is with an aggressive driving style, Figure 12 (a) shows that the proposed approach consistently outperforms traditional ACC across all cut-in gaps, achieving up to 0.9s² THT reduction. Compared with NE-MPC, it shows clear advantages in short cut-in gaps (10m and 20m), with improvements of 1.65s² and 0.66s², respectively. The proposed approach exhibits better safety performance against aggressive CVs than conservative ones, as it can identify aggressive behavior and proactively decelerate to avoid extreme and prolonged risk exposure. In contrast, interactions with conservative CVs involve right-of-way competition, which may introduce additional risk. When CV is with a neutral style (Figure 12 (c)), the proposed approach achieves the best safety performance and demonstrates better safety when facing with aggressive CVs.

*Comfort performance quantification*: The driving comfort is quantified through coefficient of variation of EV's speed as shown in Figure 13. Traditional ACC exhibits the worst comfort performance due to reactive responses to cut-ins that induce speed oscillations. Both the proposed approach and NE-MPC achieve significantly lower speed variation. The proposed approach improves comfort by 20.8% on average when interacting with conservative CVs, while being slightly less comfortable than NE-MPC for aggressive CVs due to stronger deceleration to avoid conflicts. When CV is neutral, the proposed approach is more comfortable than the baseline NE-MPC approach.

### C. *Simulation in the context of traffic*

#### 1) *Experiment design*

*Testbed*: The traffic simulation platform is PTV Vissim and Matlab/Simulink. The traffic simulation software PTV Vissim is used to build a 3 km two-lane-freeway and generate background traffic flow with different driving styles. It also provides an API named COM for Matlab/Simulink to control EV. Figure 14 explicitly depicts the structure of the testbed.

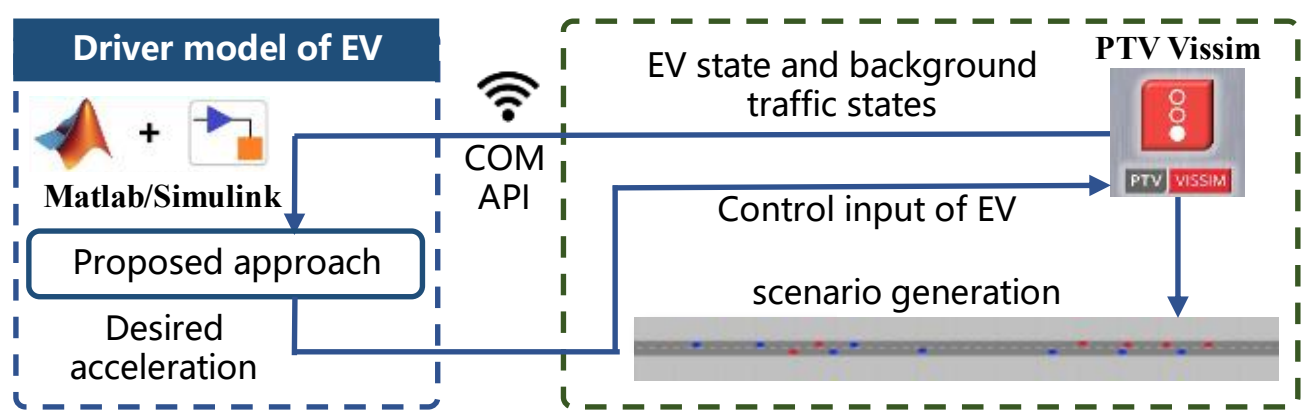

Figure 14 The structure of the testbed in traffic simulation.

*Sensitive analysis*: Sensitivity analysis is about congestion levels. Four congestion levels are tested: v/c=0.2; v/c=0.4; v/c=0.6; v/c=0.8. It is expected the proposed approach would have different performance under different congestion levels. The v/c ratio greater than or equal to 1 is not considered. That's because vehicles have difficulty executing cut-ins in oversaturated traffic conditions. Each congestion level would be simulated 30 times randomly.

*Metrics*: The following metrics are utilized for performance evaluation:

- The travel time of EV is used to measure the travel efficiency in traffic flow. It is defined as the average time spent during traveling the same distance.
- Speed standard deviation of EV is used to evaluate the robustness of the adaptive cruising performance.

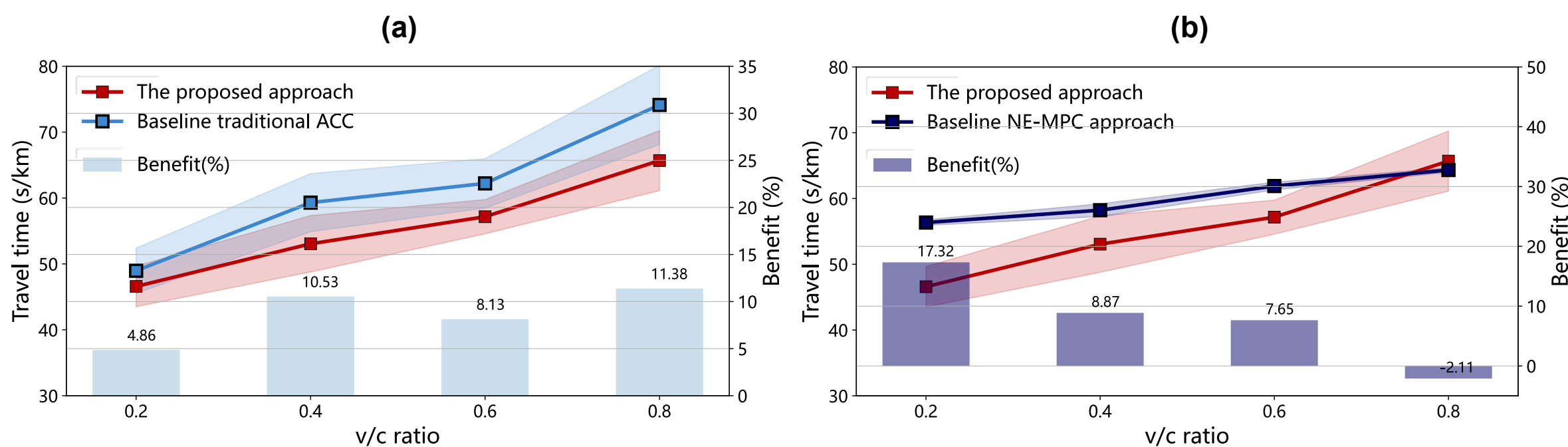

Figure 16 EV's travel time under different v/c ratios compared with (a) baseline traditional ACC; (b) baseline NE-MPC approach. The transparent ribbon denotes the 95% confidence level interval.

TABLE 2 Comparison of EV's speed standard deviation under different v/c ratios. The format is "mean value±95% confidence level".

| Control types | Speed standard deviation of EV (m/s) | | | | |
|---|---|---|---|---|---|
| | v/c=0.2 | v/c=0.4 | v/c=0.6 | v/c=0.8 | average |
| Proposed approach | **4.49±0.62** | 5.78±0.41 | **5.16±0.48** | 5.82±0.43 | **5.31±0.49** |
| Baseline traditional ACC | 4.69±0.81 | 6.60±0.54 | 5.39±0.66 | 7.01±0.62 | 5.92±0.66 |
| Baseline NE-MPC approach | 6.03±1.20 | **4.97±1.02** | 6.12±1.05 | **4.90±1.45** | 5.51±1.18 |

- Computation time is used to evaluate the computational efficiency of the proposed approach.
- Time headway distribution of EV is adopted to evaluate the safety and flexibility of the proposed approach. This research set 10s as the upper bound of vehicles' time headway.

*2) Main results*

The traffic simulation results demonstrate that the proposed approach: i) reduces travel time in traffic flow by up to 17.32%; ii) improves speed robustness by 10.3% on average; iii) adopts more flexible and competitive driving strategies; iv) supports field implementation by ensuring a 50 milliseconds computation time. Across multiple simulation runs, the improvements are statistically significant with $p < 0.05$.

*Travel efficiency in traffic flow*: The travel efficiency is evaluated by the travel time, as illustrated in Figure 16. Results show that the proposed approach achieves maximum travel time reduction by 11.38% and 17.32%, compared with baseline traditional ACC and baseline NE-MPC approach. A sensitive analysis is conducted in terms of different v/c ratios. Travel time for both the baselines and the proposed approach increases with increasing v/c ratio. It can be seen the proposed approach spent lower travel time in all v/c ratios compared with baseline traditional ACC. The benefit in the v/c ratio of 0.8 is the highest. The reason is that the proposed approach can prevent cut-ins thus avoiding being obstructed in congested traffic to obtain more travel efficiency. Compared with the baseline NE-MPC approach, the benefit gradually declines and becomes slightly negative (-2.11%) at v/c = 0.8. This indicates that the proposed approach is more effective when surrounding traffic density is low, allowing person-by-person proactive cut-in prevention. Besides, the proposed approach remains competitive with the baseline NE-MPC approach in dense traffic.

*Travel robustness in traffic flow*: The travel robustness is assessed by the speed standard deviation of EV as shown in. TABLE 2. Compared with baseline traditional ACC and baseline NE-MPC approach, results show that the speed standard deviation improvement is 10.3% and 3.6% on average. A sensitive analysis is conducted in terms of different v/c ratios. It can be seen that at all v/c ratios, the proposed approach has smaller mean speed standard deviations than baseline traditional ACC. The reason is that the proposed approach can react predictively to CVs to avoid drastic speed fluctuations. Compared with baseline NE-MPC approach, the proposed approach outperforms at v/c = 0.2 and 0.6 while the baseline outperforms at others. This indicates both approaches can quickly adapt to frequent stop-and-go patterns without enforcing anticipatory behaviors.

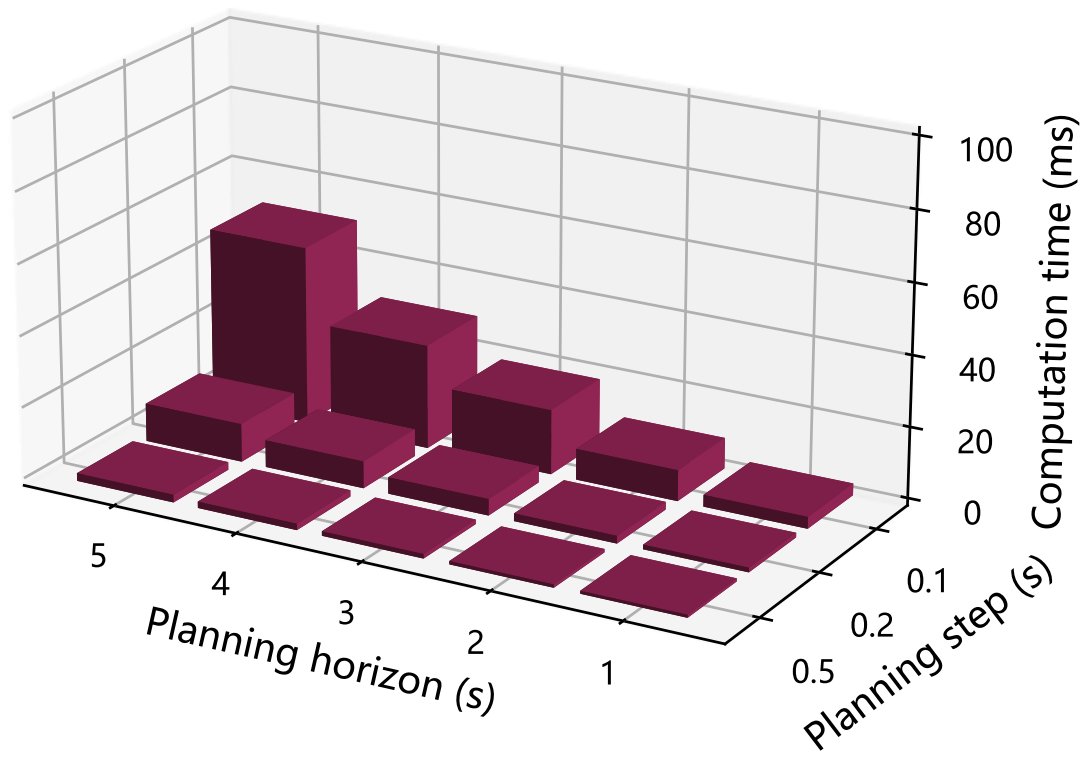

Figure 15 The computation time regarding varying settings.

*Computational efficiency*: The proposed approach supports online field implementation through computation time quantification, as illustrated in Figure 15. Results reveal that the average computation time is 9.54 milliseconds across all settings. A sensitive analysis is conducted in terms of planning horizons and planning steps. Results show that the computation time increases with increasing planning horizons and decreasing planning steps. The proposed approach can ensure a 50-millisecond computation time in nearly all cases.

*Travel flexibility in traffic flow*: The proposed approach allows for more flexible and competitive strategies, as illustrated in Figure 17. Sensitivity analysis under different v/c ratios indicates that at v/c=0.2, the proposed approach yields a larger average time headway than traditional ACC, reflecting improved safety in uncongested traffic due to effective right-of-way protection. When v/c>0.2, the proposed approach exhibits slightly smaller average time headways than traditional ACC and consistently smaller headways than NE-MPC, indicating more competitive behavior. This is because the proposed method proactively accelerates to prevent unsafe cut-ins. Furthermore, compared with traditional ACC, the time-headway distribution of the proposed approach becomes more uniform at higher v/c ratios, whereas traditional ACC remains overly centralized due to its fixed safety distance. In contrast, NE-MPC maintains an overly uniform headway distribution, suggesting limited competitiveness, while the proposed approach achieves shorter yet safe headways when necessary.

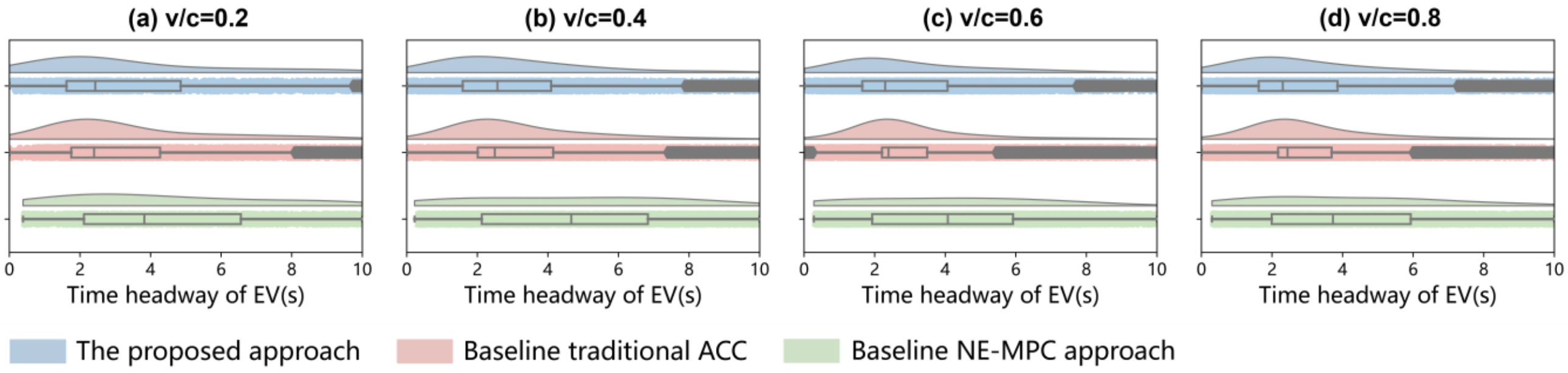

Figure 17 Comparison of EV's time headway distribution under different v/c ratios.

*Case study on the absence of CV driving style identification*: This paper further compares the performance of the proposed approach with and without the CV driving style identification through a case study. We assume that when this module is moved, the style parameters of CV are fixed as $\boldsymbol{\beta} = (1, 5, 0.1, 1, 0.1)^{\mathrm{T}}$. We extract a representative scenario from the traffic simulation and simulates the approach without the CV driving style identification module as shown in Figure 19. Comparing the position of $6^{\text{th}}$ trajectory point of EV and CV in Figure 19 (a), EV identifies CV's behavior nature and overtakes CV for cut-in prevention. By contrast in Figure 19 (b), without the CV driving style identification module, EV assumes a fixed driving style of CV and yields. This case highlights the importance of the driving style identification module.

*3) Impact analysis at the traffic flow level*

To further validate the proposed approach's impact on the traffic flow, we have raised the penetration rates of EV connecting to the cloud from 0% to 30%, for investigating what's the difference when more EVs enroll. According to the predefined penetration rate, we randomly select EVs from one lane to connect with the cloud. The v/c ratio is set to 0.8. We adopt the lane average speed (the average speed of all vehicles on the EVs' lane) as the metric.The results indicate that the lane average speed increases monotonically with the penetration rate of EVs connecting to the cloud as shown in Figure 18. When the penetration rate rises from 10% to 30%, the average speed gain grows from 0.23 km/h to 1.36 km/h, showing an increasingly pronounced improvement. This trend suggests that higher EV participation enables more effective coordination, leading to noticeable traffic-flow-level benefits.

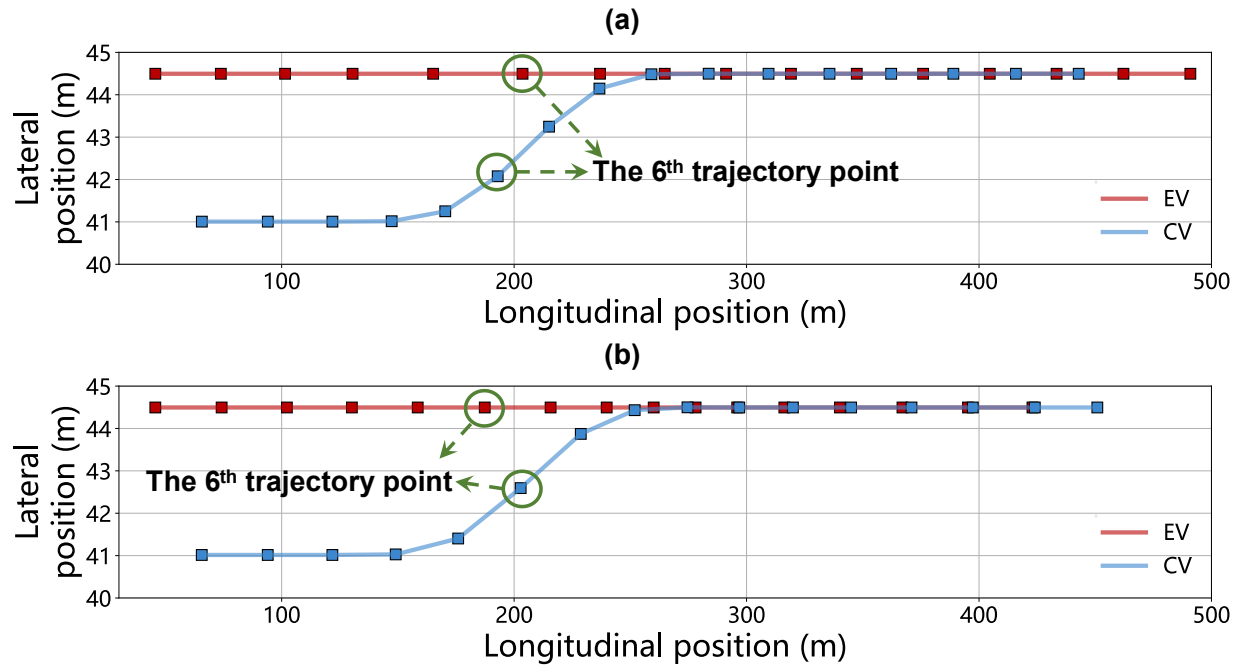

Figure 19 Sampled trajectories of (a) The proposed approach; (b) The proposed approach without CV driving style identification. The marked points indicate the positions of vehicles every 1.5 seconds.

## V. Conclusion

This research proposed an Anti-bullying Adaptive Cruise Control (AACC) approach. It is conducted based on a vehicle-cloud cooperation scheme. It bears the following features: i) with the enhanced capability of preventing "road bullying" cut-ins; ii) adaptive to various driving styles of CVs; iii) optimal but not unsafe; iv) with real-time field implementation capability. To realize the proactive right-of-way protection capability, a Game-theoretic-based MPC (GMPC) planner is designed in the cloud, and Stackelberg competition is adopted as the game rule. An online Inverse Optimal Control (IOC) based algorithm is leveraged to identify the individual driving behavior of CVs in order to enable person-by-person countermeasures. By integrating CVs' reaction functions into the system dynamics of the EV's optimal planning, the cloud could consider CVs' all possible reactions to find EV's optimal maneuver. To evaluate the proposed approach, simulation experiments were conducted.

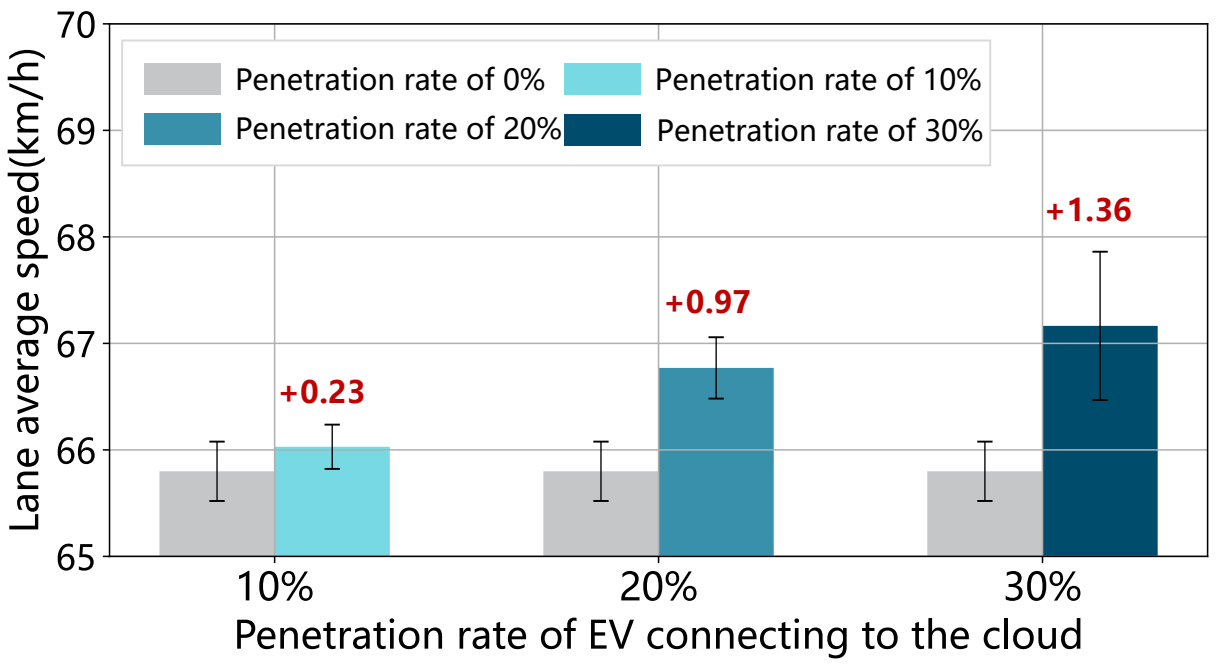

Figure 18 Lane average speed comparison when penetration rate of EV connecting to the cloud varies.

The results have shown that:

- The proposed approach can protect EV's right-of-way from "road bullying" cut-ins and is adaptive to different "road bullying" CVs' driving styles by optimizing the driving strategy in person-by-person manners.
- The proposed approach can strengthen driving safety in most cases by relieving the aggravation of potential risks.
- The proposed approach can improve comfort performance by 20.8% in average when CV with conservative driving styles.
- The proposed approach is efficient and robust against traffic congestion levels. It can reduce travel time by up to 17.32% and reduce mean speed standard deviation by 10.3% on average.
- The proposed approach can ensure less than 50 milliseconds computation time across parameter settings.
- Increasing the penetration rate of EVs connecting to the cloud has positive impact on traffic, obtaining efficiency benefit by up to 1.36 km/h.

Although the proposed approach has achieved many benefits, it still has limitations. For example, it needs to be pointed out that the proposed approach does not consider the stochasticity of human driver behavior. Further work could consider modeling $\boldsymbol{\beta}$ as a distribution for better robustness.

## APPENDIX

### A. Proof of Theorem 1

Based on Pontryagin's Minimum Principle [43], the solution to the longitudinal components of equation (8) shall satisfy the following necessary conditions:

$$\boldsymbol{\lambda}_n = \nabla_{\boldsymbol{x}_{long,n}} H_n(\boldsymbol{x}_{long,n}, a_{CV,n}, \boldsymbol{\lambda}_{n+1}, \boldsymbol{\beta}_{long}) \tag{29a}$$

$$\nabla_{a_{CV,n}} H_n(\boldsymbol{x}_{long,n}, a_{CV,n}, \boldsymbol{\lambda}_{n+1}, \boldsymbol{\beta}_{long}) = 0 \tag{29b}$$

where $\boldsymbol{\lambda}_n$ is the co-state of $\boldsymbol{x}_{long,n}$ and the Hamilton's function $H_n$ is defined as follows:

$$\begin{aligned} H_n(\boldsymbol{x}_{long,n}, a_{CV,n}, \boldsymbol{\lambda}_{n+1}, \boldsymbol{\beta}_{long}) = \boldsymbol{\beta}_{long}^{\mathrm{T}} \boldsymbol{L}_{long,n}(\boldsymbol{x}_{long,n}, a_{CV,n}) + \\ \boldsymbol{\lambda}_{n+1}^{\mathrm{T}}(\boldsymbol{A}_n \boldsymbol{x}_n + \boldsymbol{B}_n u_{EV,n} + \boldsymbol{C}_n \boldsymbol{u}_{CV,n}) \end{aligned} \tag{30}$$

In the context of the parameter identification problem, the goal is not to obtain the optimal solution via equations (29a) and (29b), but is described as: given the optimal solution (the real-time collected trajectory data of CV), a set of parameters $\boldsymbol{\beta}_{long}$ is searched for enabling the trajectory data to best satisfy the necessary condition. Substituting equation (30) into equation (29a), the following equation could be derived:

$$\begin{aligned} \boldsymbol{\lambda}_{n+1} &= (\boldsymbol{A}_n^{\mathrm{T}})^{-1}\boldsymbol{\lambda}_n - (\boldsymbol{A}_n^{\mathrm{T}})^{-1}\boldsymbol{\beta}_{long}\nabla_{\boldsymbol{x}_{long,n}}\boldsymbol{L}_{long,n} \\ &= (-(\boldsymbol{A}_n^{\mathrm{T}})^{-1}\nabla_{\boldsymbol{x}_{long,n}}\boldsymbol{L}_{long,n} \quad (\boldsymbol{A}_n^{\mathrm{T}})^{-1})\boldsymbol{\gamma}_n \end{aligned} \tag{31a}$$

$$\boldsymbol{\gamma}_n = (\boldsymbol{\beta}_{long}, \boldsymbol{\lambda}_n)^{\mathrm{T}} \tag{31b}$$

then the following recursion could be obtained:

$$\boldsymbol{\gamma}_{n+1} = \boldsymbol{K}_n \boldsymbol{\gamma}_n \tag{32a}$$

$$\boldsymbol{K}_n = \begin{pmatrix} \boldsymbol{I} & \boldsymbol{0} \\ -(\boldsymbol{A}_n^{\mathrm{T}})^{-1}\nabla_{\boldsymbol{x}_{long,n}}\boldsymbol{L}_{long,n} & (\boldsymbol{A}_n^{\mathrm{T}})^{-1} \end{pmatrix} \tag{32b}$$

$$\boldsymbol{\gamma}_{n+1} = \boldsymbol{K}_n \boldsymbol{K}_{n-1} \boldsymbol{K}_{n-2} \ldots \boldsymbol{K}_0 \boldsymbol{\gamma}_0 = \boldsymbol{\mathcal{K}}_n \boldsymbol{\gamma}_0 \tag{32c}$$

$$\boldsymbol{\gamma}_0 = (\boldsymbol{\beta}_{long}, \boldsymbol{\lambda}_0)^{\mathrm{T}} \tag{32d}$$

where $\boldsymbol{I}$ denotes identity matrix and $\boldsymbol{0}$ denotes null matrix. Substituting equation (30) into equation (29b), the following equation could be derived:

$$(\nabla_{a_{CV,n}}\boldsymbol{L}_{long,n} \quad \boldsymbol{C}_{n+1}^{\mathrm{T}})\boldsymbol{\gamma}_{n+1} = \boldsymbol{\mathcal{L}}_n \boldsymbol{\gamma}_{n+1} = 0 \tag{33}$$

substituting equation (32c) into (33), the following linear equation could be derived:

$$\boldsymbol{\mathcal{L}}_n \boldsymbol{\mathcal{K}}_n \boldsymbol{\gamma}_0 = 0 \tag{34}$$

Thus, the longitudinal driving style parameters $\boldsymbol{\beta}_{long}$ of CV can be identified by satisfying the equation (34).

### B. Proof of Theorem 2

This research assumes that the reaction of CV follows an optimal control scheme and can be obtained by solving the following optimal control problem. The objective is transformed from equation (12):

$$\min_{\boldsymbol{u}_{CV,n}|_{n=0}^{N-1}} \left( \sum_{n=0}^{N-1} \frac{1}{2} \begin{bmatrix} (\boldsymbol{x}_n - \boldsymbol{x}_{des}^{CV})^{\mathrm{T}} \boldsymbol{Q}_{CV} (\boldsymbol{x}_n - \boldsymbol{x}_{des}^{CV}) + \\ \boldsymbol{u}_{CV,n}^{\mathrm{T}} \boldsymbol{\beta}_d \boldsymbol{u}_{CV,n} \end{bmatrix} \right) \tag{35}$$

where,

$$\boldsymbol{Q}_{CV} = diag(\beta_1, 0, \beta_2, \beta_4, \beta_5) \tag{36a}$$

$$\boldsymbol{x}_{des}^{CV} = (\Delta x_{des}^{CV}, 0, v_{des}^{CV}, y_{des}^{CV}, \psi_{des}^{CV})^{\mathrm{T}} \tag{36b}$$

$$\boldsymbol{x}_n = \boldsymbol{A}_{n-1}\boldsymbol{x}_{n-1} + \boldsymbol{B}_{n-1} u_{EV,n-1} + \boldsymbol{C}_{n-1}\boldsymbol{u}_{CV,n-1} \tag{36c}$$

$$\boldsymbol{\beta}_d = diag(\beta_3, 0) \tag{36d}$$

It can be found that $\boldsymbol{Q}_{CV}$ and $\boldsymbol{\beta}_d$ are from the identified driving style $\boldsymbol{\beta}$. This optimization problem could be solved by the dynamic programming algorithm. The cost-to-go function is defined as:

$$V(\boldsymbol{x}_i) = \min_{\boldsymbol{u}_{CV,n}|_{n=i}^{N-1}} \left( \begin{matrix} \displaystyle\sum_{n=i}^{N-1} \frac{1}{2} \begin{bmatrix} (\boldsymbol{x}_n - \boldsymbol{x}_{des}^{CV})^{\mathrm{T}} \boldsymbol{Q}_{CV} (\boldsymbol{x}_n - \boldsymbol{x}_{des}^{CV}) + \\ \boldsymbol{u}_{CV,n}^{\mathrm{T}} \boldsymbol{\beta}_d \boldsymbol{u}_{CV,n} \end{bmatrix} \\ +V(\boldsymbol{x}_N) \end{matrix} \right) \tag{37a}$$

$$\begin{aligned} V(\boldsymbol{x}_N) &= \frac{1}{2}(\boldsymbol{x}_N - \boldsymbol{x}_{des}^{CV})^{\mathrm{T}} \boldsymbol{M}_N (\boldsymbol{x}_N - \boldsymbol{x}_{des}^{CV}) + \boldsymbol{O}_N^{\mathrm{T}} \boldsymbol{x}_N \\ V(\boldsymbol{x}_N) &= 0 \end{aligned} \tag{37b}$$

where $\boldsymbol{O}_N = (0,0,0,0,0)^{\mathrm{T}}$, $\boldsymbol{M}_N = diag(0,0,0,0,0)$. Based on Bellman's optimality principle, the solution to equation (35) can be achieved via iteratively optimizing this cost-to-go function $V(\boldsymbol{x}_i)$ in reverse order. At time step $N-1$, the cost-to-go function can be specific as:

$$V(\boldsymbol{x}_{N-1}) = \min_{\boldsymbol{u}_{CV,N-1}} \frac{1}{2} \begin{bmatrix} (\boldsymbol{x}_{N-1} - \boldsymbol{x}_{des}^{CV})^{\mathrm{T}} \boldsymbol{Q}_{CV} (\boldsymbol{x}_{N-1} - \boldsymbol{x}_{des}^{CV}) \\ +\boldsymbol{u}_{CV,N-1}^{\mathrm{T}} \boldsymbol{\beta}_d \boldsymbol{u}_{CV,N-1} \end{bmatrix} + V(\boldsymbol{x}_N) \tag{38}$$

Substituting the equation (36c) into the equation (38) and minimize the cost-to-go function based on the algorithm developed by this research team [26], and then the optimal reaction function of CV can be derived as equation (19). It is easy to find that when time step is $N-2, N-3, \ldots$, the reaction function is consistent.

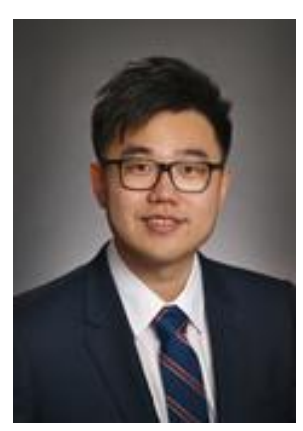

**Jia Hu** (Senior Member, IEEE) is currently working as a Zhongte Distinguished Chair of Cooperative Automation with the College of Transportation Engineering, Tongji University. Before joining Tongji University, he was a Research Associate with the Federal Highway Administration (FHWA), USA. He is an Editorial Board Member of the *Journal of Intelligent Transportation Systems* and the *International Journal of Transportation Science and Technology*. He is a member of TRB (a Division of the National Academies) Vehicle Highway Automation Committee, the Freeway Operations Committee, Simulation subcommittee of Traffic Signal Systems Committee, and the Advanced Technologies Committee of the ASCE Transportation and Development Institute. He is the Chair of the Vehicle Automation and Connectivity Committee of the World Transport Convention. He is an Associate Editor of the American Society of Civil Engineers *Journal of Transportation Engineering* and IEEE OPEN JOURNAL OF INTELLIGENT TRANSPORTATION SYSTEMS.

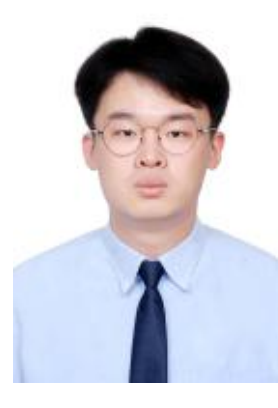

**Zhexi Lian** was born in Shanxi, China. He received the bachelor's degree in traffic engineering from College of Transportation, Tongji University, Shanghai, China, in 2023. He is currently pursuing the Ph.D. degree with Key Laboratory of Road and Traffic Engineering of the Ministry of Education, Tongji University. His main research interests include data-driven control and end-to-end autonomous driving.

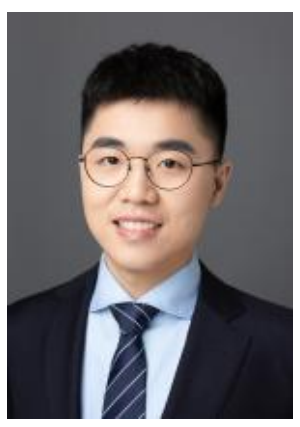

**Haoran Wang** (Member, IEEE) received the bachelor's degree in transportation engineering from Tongji University, Shanghai, China, in 2017, and the Ph.D. degree from Tongji University in 2022. He is currently a Postdoctoral Researcher with the College of Transportation Engineering, Tongji University. He is a researcher on vehicle engineering, majoring in intelligent vehicle control and cooperative automation. Dr. Wang served the IEEE TRANSACTIONS ON INTELLIGENT VEHICLES, IEEE TRANSACTIONS ON INTELLIGENT TRANSPORTATION SYSTEMS, *Journal of Intelligent Transportation Systems*, and IET *Intelligent Transport Systems* as peer reviewers with a good reputation.

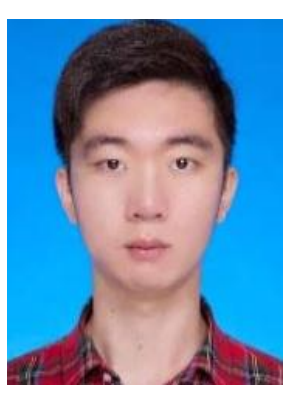

**Zihan Zhang** was born in Shanghai, China. He received the B.S. degree in traffic engineering from Dalian Maritime University, Dalian, China, in 2018, and received the Ph.D degree from Tongji University in 2023. He is currently the engineer with Shanghai Motor Vehicle Inspection Certification and Tech Innovation Center Co., LTD, Shanghai. His research interests include connected and automated vehicle, ITS, optimal control theory, eco-driving, and human-like decision-making and control.

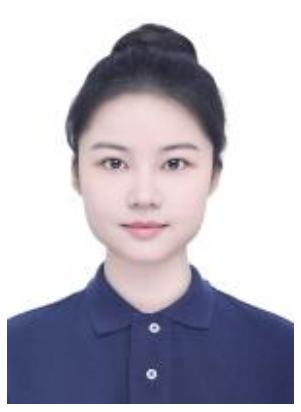

**Ruoxi Qian** is currently pursuing the bachelor's degree in Mathematics with Economics with University College London, London, England. Her main research interests include data science application in autonomous driving systems.

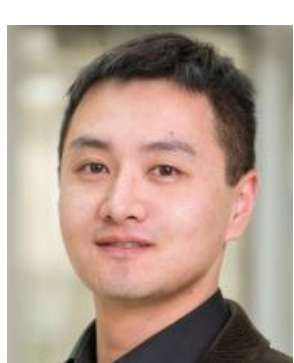

**Duo Li** (Senior Member, IEEE) received the B.E. degree in civil engineering (transportation) from the Huazhong University of Science and Technology in 2010, the M.S. degree from The University of Queensland in 2011, and the Ph.D. degree from the University of Auckland in 2015. Since 2022, he has been a Senior Lecturer with the Department of Engineering, Nottingham Trent University. Before this, he had several academic and research positions including a Research Associate with the University of Cambridge, a Humboldt Research Fellow with the German Aerospace Center (DLR), and an academic positions with Chang'an University. His research interests include intelligent transport system (ITS) optimization, microscopic and macroscopic transport modeling and simulation, and data-driven traffic analytics and forecasting.

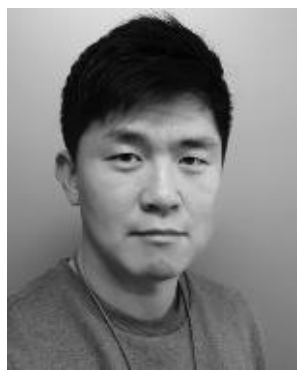

**Jaehyun (Jason) So** received the B.S. degree and the M.S. degree in transportation engineering from Ajou University, South Korea, in 2006 and 2008, respectively, and the Ph.D. degree in civil and environmental engineering from the University of Virginia, VA, USA, in 2013. He is currently an Assistant Professor with Department of Transportation System Engineering at Ajou University. His research interests include traffic operations, safety, and connected-automated vehicles.

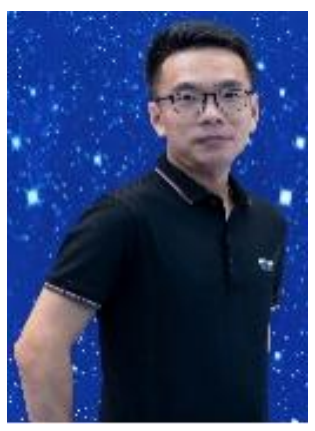

**Junnian Zheng** was born in Shanghai, China. He holds B.S. degree in mechanical engineering from Shanghai Jiao Tong University, and M.S. and Ph.D. degrees in mechanical engineering from Texas A&M University. He is currently the director of innovation and advanced engineering at Hyperview Mobility (Shanghai). His research interests include ADAS and autonomous driving system for passenger and commercial vehicles, embodied AI, and large language models for self-driving.